\documentclass[structabstract]{aa}  
\usepackage{graphicx}
\usepackage{subfig}
\usepackage{nth}
\usepackage{longtable}
\usepackage{color}

\usepackage{txfonts}
\usepackage{natbib}
\bibpunct{(}{)}{,}{a}{}{,}
\newcommand{\ms}{\mbox{${\rm m\,s}^{-1}$}}

\newcommand{\Msolar}{\mbox{${M}_{\sun}$}}
\newcommand{\Rsolar}{\mbox{${R}_{\sun}$}}

\newcommand{\Rstellar}{\mbox{${R}_{\ast}$}}

\newcommand{\Mjup}{\mbox{${M}_{J}$}}

\newcommand{\rhosun}{\mbox{$\rho_{\sun}$}}
\newcommand{\Rjup}{\mbox{${R}_{J}$}}
\newcommand{\rhojup}{\mbox{$\rho_{J}$}}
\newcommand{\mym}{\mbox{$\muup$}\rm m}

\newcommand\T{\rule{0pt}{2.2ex}}

\begin{document}

   \title{FORS2 observes a multi-epoch transmission spectrum of the hot Saturn-mass exoplanet WASP-49b\thanks{Based on photometric observations made with FORS2 on the ESO VLT/UT1 (Prog. ID 090.C-0758),
 EulerCam on the Euler-Swiss telescope and the Belgian TRAPPIST telescope.}$^{,}$\thanks{The photometric time series data in this work are only available in electronic form
at the CDS via anonymous ftp to cdsarc.u-strasbg.fr (130.79.128.5) or via http://cdsweb.u-strasbg.fr/cgi-bin/qcat?J/A+A/}}
   
   \author{M.~Lendl
          \inst{1,2,3}
          \and
          L.~Delrez
          \inst{2}
          \and
          M.~Gillon
          \inst{2}
          \and
          N.~Madhusudhan
          \inst{4}
          \and
          E.~Jehin
          \inst{2}
	  \and
          D.~Queloz
          \inst{3,5} 
          \and
          D.R. Anderson\inst{6}
          \and 
          B.-O.~Demory\inst{5}
          \and
          C. Hellier\inst{6}
}

   \institute{Space Research Institute, Austrian Academy of Sciences, Schmiedlstr. 6, 8042 Graz, Austria, 
              \email{monika.lendl@oeaw.ac.at}
         \and
             Universit\'e de Li\`ege, All\'ee du 6 ao\^ut 17, Sart Tilman, Li\`ege 1, Belgium
         \and
             Observatoire de Gen\`eve, Universit\'e de Gen\`eve, Chemin des maillettes 51, 1290 Sauverny, Switzerland,
         \and 
             University of Cambridge, Madingley Road, Cambridge CB3 0HA, UK
         \and 
             Cavendish Laboratory, J J Thomson Avenue, Cambridge, CB3 0HE, UK
         \and
             Astrophysics Group, Keele University, Staffordshire, ST5 5BG, UK
           }
  \date{}

 
  \abstract
   {
Transmission spectroscopy has proven to be a useful tool for the study of exoplanet atmospheres, because the absorption and scattering signatures 
of the atmosphere manifest themselves as variations in the planetary transit depth. Several planets have been studied with this
technique, leading to the detection of a small number of elements and molecules (Na, K, $\mathrm{H_2O}$), but also revealing that 
many planets show flat transmission spectra consistent with the presence of opaque high-altitude clouds.
}
   {
We apply this technique to the $M_P=0.40$~$\Mjup$, $R_p=1.20$~$\Rjup$, $P=2.78$~d planet WASP-49b, aiming to characterize its 
transmission spectrum between 0.73 and 1~{\mym} and search for the features of K and $\mathrm{H_2O}$. 
Owing to its density and temperature, the planet is predicted to possess an extended atmosphere and 
is thus a good target for transmission spectroscopy.
}
   {
Three transits of WASP-49b have been observed with the FORS2 instrument installed at the VLT/UT1 telescope at the ESO Paranal site. 
We used FORS2 in MXU mode with grism GRIS\_600z, producing simultaneous multiwavelength transit light curves throughout 
the i' and z' bands. We combined these data with independent broadband photometry from the Euler and TRAPPIST telescopes to obtain
a good measurement of the transit shape. Strong correlated noise structures are present in the FORS2 light curves, which are due to 
rotating flat-field structures that are introduced by inhomogeneities of the linear atmospheric dispersion corrector's transparency.
We accounted for these structures by constructing common noise models from the residuals of light curves bearing the same noise structures and
used them together with simple parametric models to infer the transmission spectrum.
}
   {
We present three independent transmission spectra of WASP-49b between 0.73 and 1.02~{\mym}, as well as a  
transmission spectrum between 0.65 and 1.02~{\mym} from the combined analysis of FORS2 and broadband data. 
The results obtained from the three individual epochs agree well. The transmission spectrum of WASP-49b is 
best fit by atmospheric models containing a cloud deck at pressure levels of 1~mbar or lower.
}
   {}
   \maketitle
%

\section{Introduction}

The study of transiting planets has become one of the main avenues for characterizing exoplanets. 
Transit light curves are observed while the planet passes between its host star and an Earth-based observer,
and many pieces of information on the planetary system are contained in them. Most prominently, the planetary radius and, in 
conjunction with a mass estimate, the planetary density are measured. 
The atmospheric properties of transiting planets are accessible to study mainly through transmission
and emission spectroscopy, that is, through multiwavelength observations of transits and occultations
(for a summary, see, e.g., \citealt{Winn11}).

Transmission spectroscopy (e.g. \citealt{Seager00, Charbonneau02}) is sensitive to the absorption features imprinted by the 
planetary atmosphere on the stellar light that passes through it during transit. In this configuration 
the planetary day-night terminator region is probed. The angle between the planetary surface and the 
incident stellar radiation causes the outer atmospheric layers to have a higher weight for these observations than for the emissive case. 

On the observational side, a limitation to transmission spectroscopy is given by stellar activity. Non-occulted spots slightly 
affect the measured transit depth \citep[e.g.,][]{McCullough2014}. These effects are largely eliminated for 
inactive planet hosting stars and can be further decreased by carrying out simultaneous observations in the 
available wavelength channels. Spectrophotometry consists of spectrally
dispersing the light of target and reference stars and then binning the spectra to a lower resolution and performing
relative photometry on the summed stellar flux in these bins. In this way, simultaneous multiwavelength observations
of transits can be obtained. Initial results stem from space-based observatories \citep{Barman07,Knutson07a,Desert08}, but more recently, 
this technique has also been used in ground-based instruments where capabilities of obtaining spectra of multiple objects allow
using comparison stars \citep{Bean10,Bean11,Sing11b}.

From high-resolution spectra, several absorption features, in particular that of Na \citep{Charbonneau02,Redfield08},
have been identified in the optical transmission spectra of giant planets. 
Initial near-IR detections based on HST/NICMOS data \citep[e.g.,][]{Swain08} have given rise to debate because independent 
analyses have yielded different results \citep{Sing09b,Gibson11,Crouzet12}. More recently, HST/WFC3 has been used on a 
few hot Jupiters, where absorption features of $\mathrm{H_2O}$ could be identified \citep[e.g.,][]{Deming13,Huitson13}.
Compared to theoretical predictions \citep[e.g.,][]{Seager00}, these signatures have been less pronounced than expected,
indicating that an additional, grayer, opacity source is present in the planetary atmospheres. 
This picture is supported by largely flat optical transmission spectra observed for several hot Jupiter planets
\citep[e.g.,][]{Pont08,Bean13,Gibson13}. If the slope of the transmission spectrum is measured across a 
broad wavelength region, the haze components can be revealed thanks to their Rayleigh scattering signature \citep[e.g.,][]{Lecavelier08a}.

The subject of this paper, WASP-49b, \citep{Lendl12} is a hot Saturn discovered by the WASP survey \citep{Pollacco06}.
WASP has been carrying out a search for hot Jupiters orbiting bright ($\mathrm{m_V<13}$) stars all across the sky.
\citet{Lendl12} measured a mass of 0.38~{\Mjup} and a radius of 1.12~{\Rjup} for WASP-49b, which is orbiting a G6~V star every 2.78 days. 
Given its low density (0.27~{\rhojup}) and short orbital period, the planet possesses an extended atmosphere and is thus a favorable target for transmission 
spectroscopy.

We here present transit observations of WASP-49b, obtained with the FORS2 instrument installed at the VLT/UT1.
Observations of three transits of WASP-49b were obtained in multi-object spectroscopy (MXU) mode, covering wavelengths 
from 0.7~{\mym} to 1.02~{\mym}. These data are supplemented by additional transit observations from the EulerCam \citep{Lendl12}
and TRAPPIST \citep{Gillon11a,Jehin11} instruments. 
In Sect. \ref{sec:W49dat}, we report details of the observations and the data reduction, and in Sect. \ref{sec:W49mod}
we describe the modeling process. The resulting transmission spectrum is shown and interpreted in Sect. \ref{sec:res},
before we conclude in Sect. \ref{sec:W49con}.

\section{Observations and data reduction}
\label{sec:W49dat}

\subsection{FORS2 spectrophotometry}
\subsubsection{Observations}
\begin{figure}[t!]
 \centering
 \includegraphics[width=\linewidth]{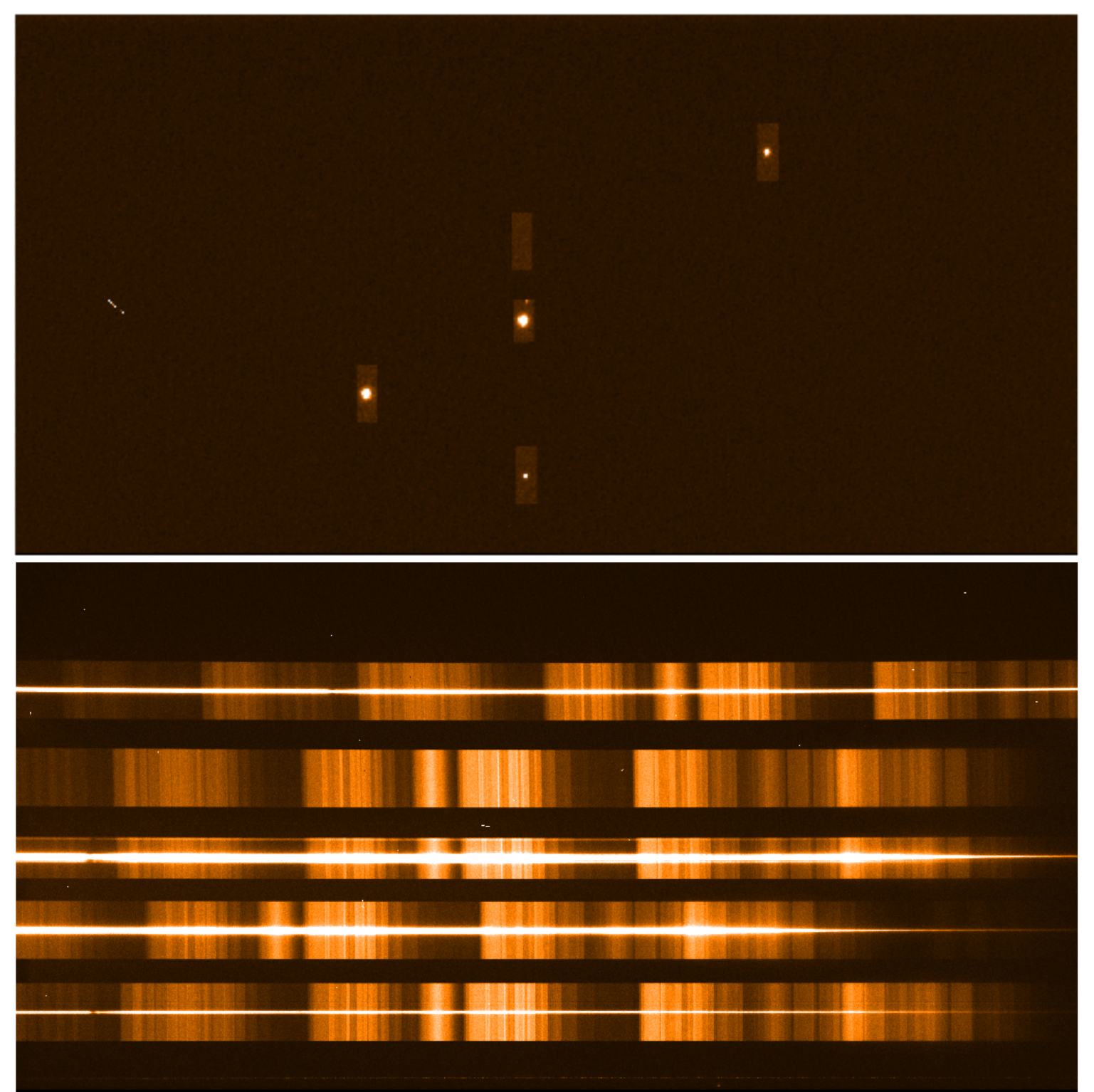}
 \caption{\label{fig:W49MXU}Top: An acquisition image for the WASP-49 FORS2 observations. The mask with five slits is visible, four of the slits
are placed on stars, one is placed on the sky. Bottom: The same field and instrument setup, but with the dispersive element in the optical path.
}
\end{figure}

Three transits of WASP-49b were observed with FORS2 \citep{Appenzeller98} at the VLT/UT1 
during the nights of 05 December 2012, 14 January 2013 and 07 February 2013, under program 090.C-0758.
The instrument was used in MXU mode, which allows performing ($R \sim 1000$) multi-object spectroscopy with the help of 
laser-cut masks made specifically for the observed field. We used wide 10 by 28 arcsec (in one case 10 by 20 arcsec) slits 
to select WASP-49 and three reference stars, and used grism GRIS\_600z for the dispersion together with order sorting filter OG590.
The large slit widths are needed to avoid flux losses during variations in seeing or pointing.
The resulting wavelength range is $738-1026$~nm for WASP-49. The wavelength range of the reference stars is slightly different owing 
to their position on the detector and thus the displacement of the spectra on the chip. The positioning of the target and reference stars is
shown in Fig. \ref{fig:W49MXU} and an example of the obtained spectra in Fig. \ref{fig:spec}.
For the wavelength calibration we used a HeArNe lamp spectrum, but narrower (0.5~arcsec) slits were used to provide 
well-defined unsaturated emission lines to match with the database. 

The linear atmospheric dispersion corrector (LADC) of FORS2 proved to impose a major limitation to the instrument's photometric performance
\citep{Moehler10} throughout several years until its upgrade in 2014 \citep{Sedaghati15}. The data treated here fall into this period. 
The LADC is composed of two prisms whose separation 
is changed with airmass to compensate for the image dispersion caused by the atmosphere. These prisms show structures of uneven transmission. 
As the LADC is located in the optical path above the image derotator, this creates structures 
that rotate across the field of view during long observing sequences. To reduce noise stemming from the LADC, the LADC prism separation was set to a 
constant value throughout each of our transit observations. For the first two nights this value was set by the previous instrument configuration, 
that is, 155.0~mm for 05 December 2012, and 898.1~mm for 14 January 2013. As we observed that the long-term correlated noise in the photometry was reduced 
for the smaller prism separation, we set the LADC to a minimum separation of 30~mm for our third (14 February 2013) observation. 

The weather during the first two transits was good, with stable seeing around 0.9~arcsec on 05 December 2012 and seeing varying between 0.8 and 1.5 arcsec 
on 14 January 2013. The data obtained on 07 February 2013 were affected by variable and partially unfavorable seeing, between 1.0 and 2.5~arcsec. 
The exposure times used were 30~s and 25~s for the first observation, and 20~s for the other observations. The third observation was interrupted 
by a technical malfunction before the beginning of the transit.

\begin{figure}
 \centering
 \includegraphics[width=0.5\linewidth]{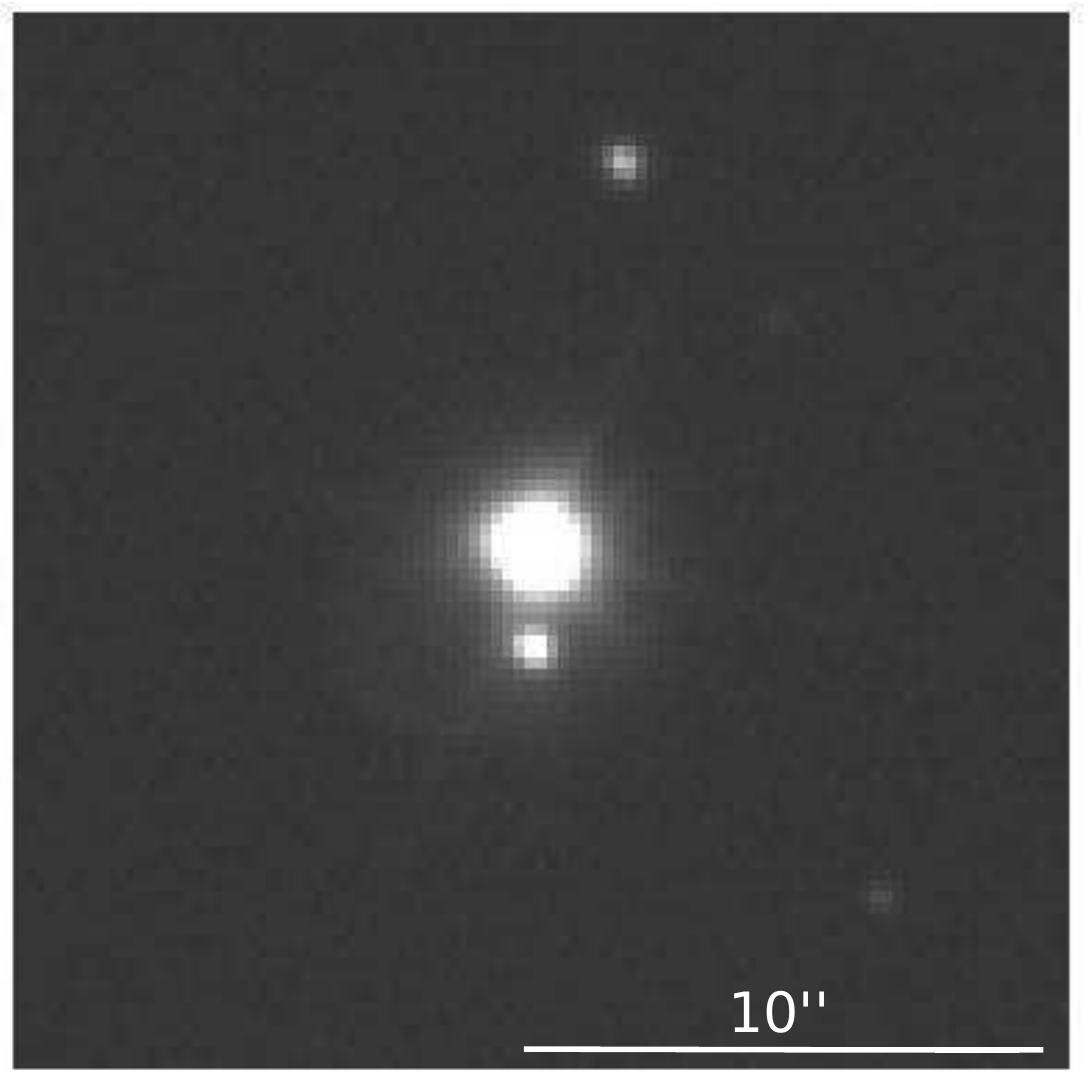}
 \caption{\label{fig:cont}WASP-49b and the nearby star, as seen in the pre-imaging with FORS2. North is up and east is left.}
\end{figure}

During the pre-imaging of the target field, we discovered a faint star ($\mathrm{\Delta mag_{z}= 4.303 \pm 0.12}$) located
2.3 arcsec south of WASP-49 (see Fig. \ref{fig:cont}). This star was blended with WASP-49 in previous observations. 

\subsubsection{Data reduction}

\begin{figure}[t]
\includegraphics[width=\linewidth]{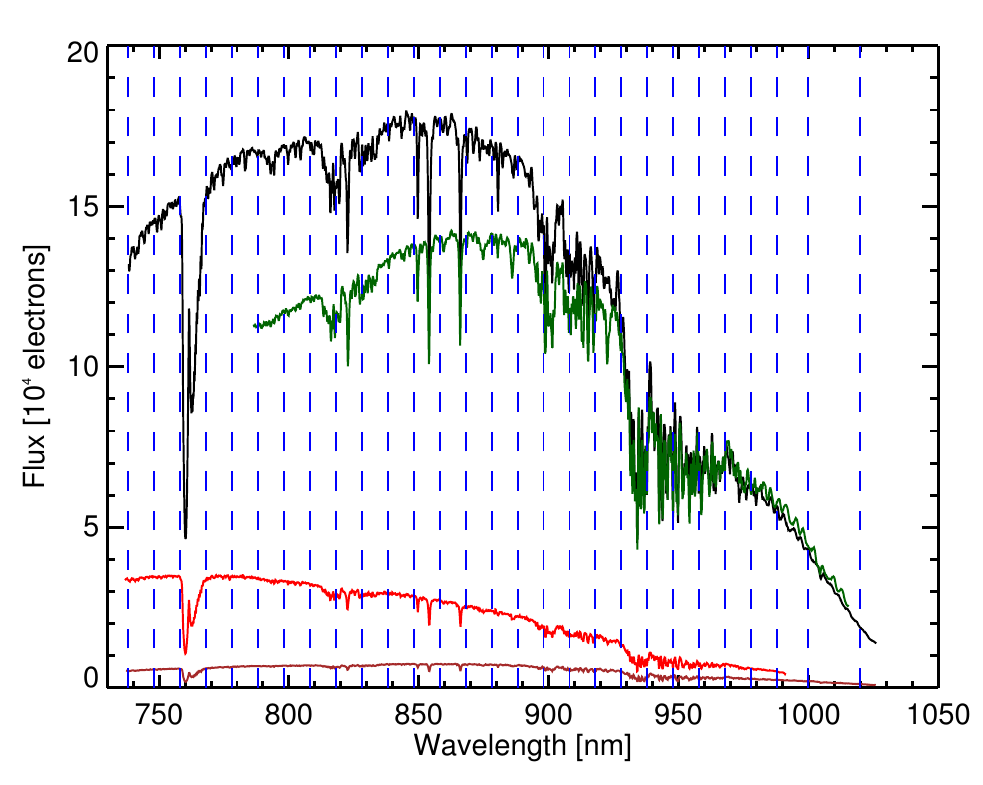}
\caption{\label{fig:spec}Example of the spectra of WASP-49 (top, black), and the three reference stars (green, red, brown). The blue dashed lines
indicate the bin sizes used for spectrophotometry.  Note that only two faint reference stars are available for the five shortest-wavelength bins.}
\end{figure}

\begin{figure}
 \includegraphics[width=\linewidth]{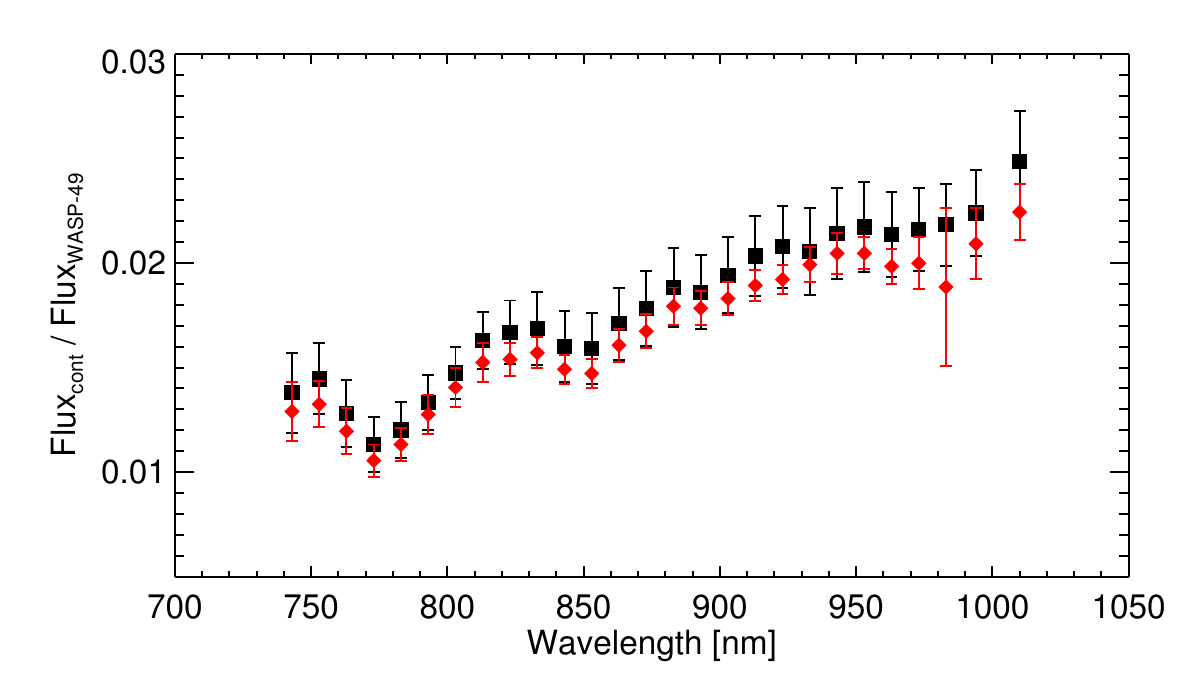}
 \caption{\label{fig:W49Con}Relative flux of the nearby star contaminating the WASP-49 spectra, shown as a function of wavelength. 
The red diamonds show the contamination estimated from photometry after subtracting the target PSF, and black squares denote the 
results from comparison of the fitted PSF peaks. The slope of the measurements indicates an object redder than WASP-49.
}
\end{figure}

The standard ESO pipeline was used to produce the master calibration frames, and to determine a wavelength solution in form 
of a third-order order polynomial based on the lamp frames. The wavelength solution was later refined by matching prominent absorption lines in the mean stellar
spectra. To extract of spectrophotometric measurements, we proceeded as follows. For each pixel, the PSF in the spatial direction was determined 
iteratively by fitting Moffat functions \citep{Moffat69} using the \texttt{mpfit} routines \citep{Markwardt09}. 
Outliers (mostly cosmic ray hits) were rejected at this step and then replaced by the values of the PSF fit at this point. 
The sky background was individually measured for each spectral pixel by fitting a first-order polynomial in the spatial direction selecting only 
regions well outside the stellar PSF. The sky contribution was then removed by subtracting this fit from all spatial pixels.
By assuming a varying sky value for each spectral pixel, we compensated for slight variations in the background that are due to bends in the 
spectra with respect to the CCD pixel grid. 

The 1D spectra were extracted for each spectral pixel, by summing the flux in several windows of different 
widths centered on the PSF peak. At this point, data affected by saturation of the detector during the 05 December 2012 observation were identified and 
removed from further analysis.
To measure the amount of contamination introduced by the nearby star, we subtracted the stellar PSF from the 2D spectrum of WASP-49 and then measured the 
contaminant's flux that fell inside each of the extraction windows. 
For a second estimate of the target to contaminant flux ratio, the PSF of the contaminant was fitted after the removal of the target PSF, and the peak values were compared. 
The resulting values averaged for all three transits and for each spectral bin are shown in Fig. \ref{fig:W49Con}.
After we extracted the spectra of all exposures, we removed outliers once more, this time based on the temporal domain. For each spectral pixel,
the extracted flux values were fit with a fourth-order polynomial with respect to time, outliers were identified and replaced by the values of the fit 
at the same position. 

Next we binned the final spectra in 27 wavelength bins, 25 of which had a width of 10~nm, only the two longest-wavelength bins measured 12~nm and 20~nm.
The location of these bins with respect to the spectra of target and reference stars is shown in Fig. \ref{fig:spec}. The five shortest-wavelength
bins and the longest-wavelength bin are not covered by all reference stars. Relative photometric light curves were created for each extraction window from 
the binned spectra. All combinations of reference stars were tested; the best light curves were obtained using all references available in each wavelength bin. 

\begin{figure*}
 \includegraphics[width=\linewidth]{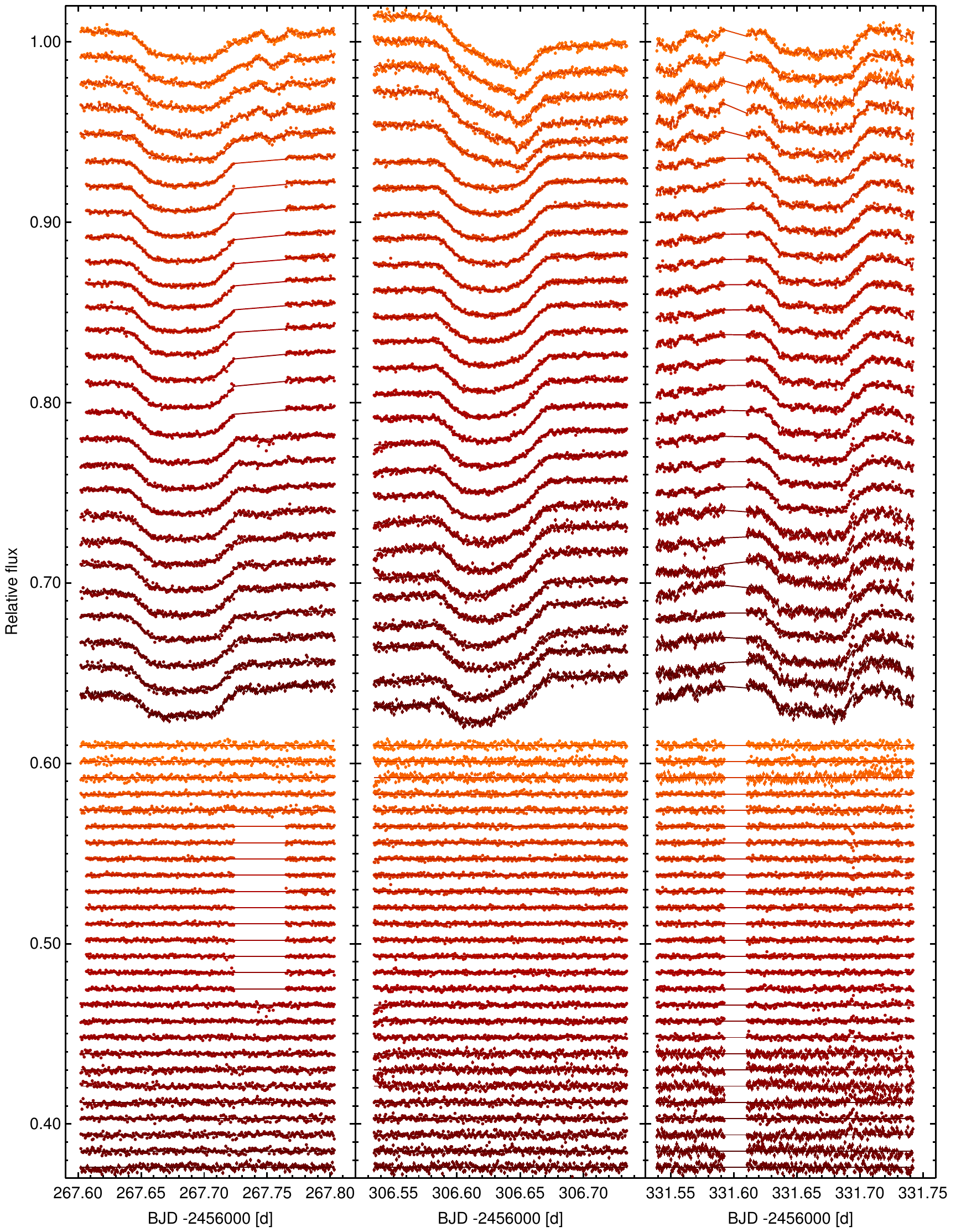}
 \caption{\label{fig:W49LC}Spectro-photometric FORS2 transit light curves of WASP-49. The wavelength of the individual light curves increases 
downward, and the bin centers are 743, 753, 763, 773, 783, 793, 803, 813, 823, 833, 843, 853, 863, 873, 883, 893, 903, 913, 923, 933,
943, 953, 963, 973, 983, 994, and 1010~nm. The residuals are shown below the data. }
\end{figure*}

For the further analysis, the light curves obtained from large extraction windows were used: 32 pixels for the transits on 05 December 2012 and 14 January 2013, 
and 36 pixels for the transit on 07 February 2013. This way, the contaminating star was contained in the aperture and its contribution to the light curve kept as stable
as possible. The resulting light curves are displayed in Fig. \ref{fig:W49LC}.

\subsubsection{FORS2 data of 05 December 2012}

The FORS2 observations of 5 December 2012 were carried out throughout the transit using an exposure time of 30~s. The peak region of the target 
spectrum exceeded the nonlinear range of the detector from $\sim$05:15~UT on, until the exposure time was adapted down to 25~s at 06:14~UT. 
Points affected by this episode of saturation were identified during the spectral extraction and removed from further analysis.

The $ \lambda < 788$~nm light curves show a very particular wave-like pattern around meridian passage. These are the same light curves that were
created using only two reference stars. This effect arises because the spatial inhomogeneities of 
the LADC transparency creates differences in the light curves of the comparison stars.

\subsubsection{FORS2 data of 14 January 2013}

Again, the $\lambda < 788$~nm light curves obtained with FORS2 on 14 January 2013 show a wave-like pattern around meridian passage, probably for the same reason as 
for the 5 December 2012 data. At the same time, the overall light curve shapes are more strongly affected, with large-scale tilts that vary in shape and 
amplitude. This is probably related to the fact that the LADC separation was large for this observation, 898.1~mm as opposed to 155.0~mm for the 
05 December 2012 observation.

\subsubsection{FORS2 data of 07 February 2013}

The FORS2 data taken on 7 February 2013 were affected by less favorable conditions, than the other transit observations, in particular, by bad seeing. 
The observations were interrupted because of a technical problem at 02:08~UT, but were resumed at 02:29~UT, \textasciitilde20 minutes before the start
of the transit. The data obtained before the interruption show large variations, related to unfavorable observing conditions and the passage of meridian 
and hence fast LADC movement. Some light curves show an unexplained short-term increase in flux during egress, which probably is of instrumental origin.

\subsection{EulerCam and TRAPPIST photometry}

\begin{figure}
\sidecaption
\includegraphics[width=\linewidth]{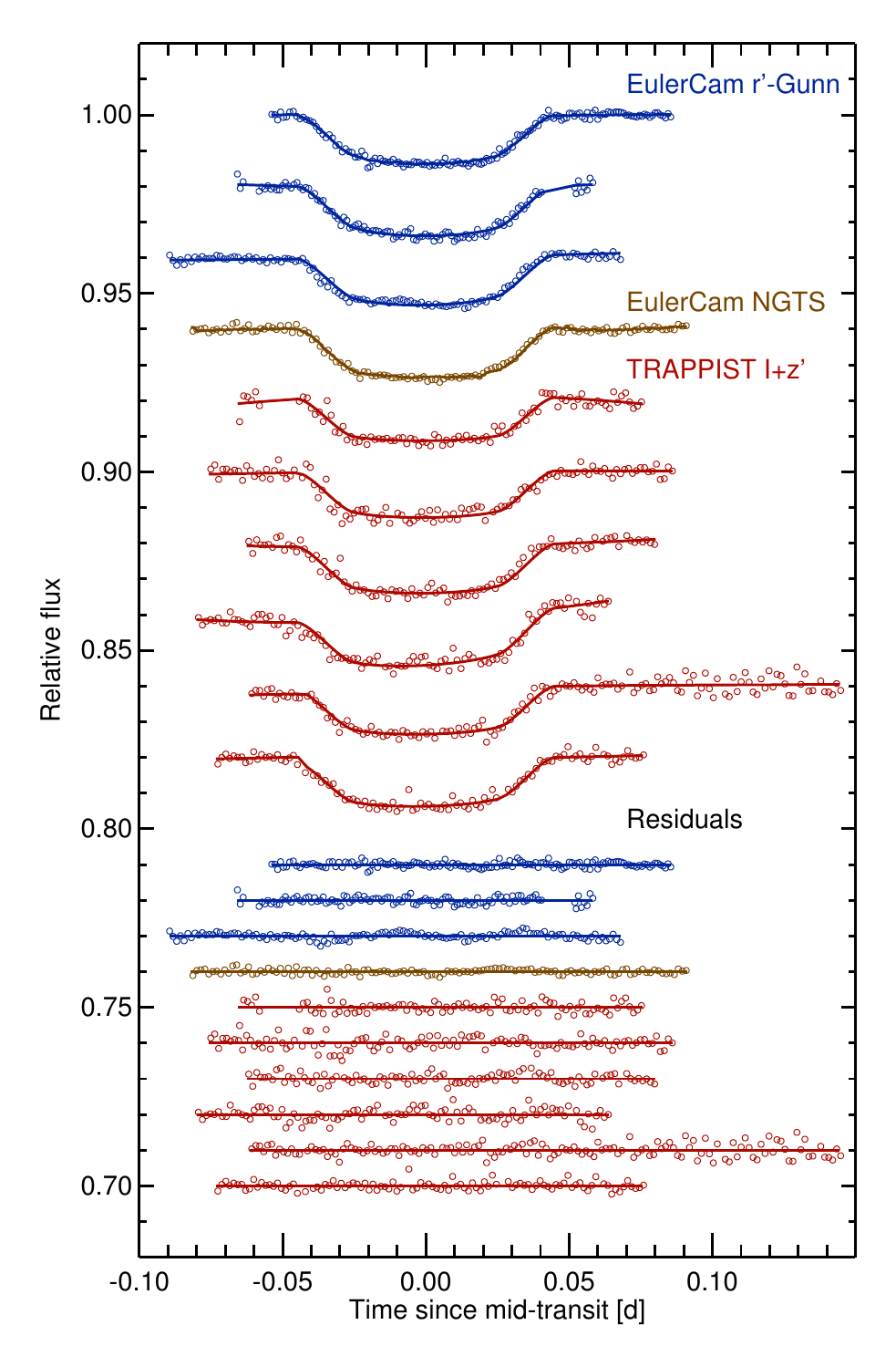}
 \caption{\label{fig:W49TE}WASP-49 transit light curves from EulerCam and TRAPPIST included in the analysis. The instrument and filters are 
color coded and are (from top to bottom) EulerCam using an r'-Gunn filter (top three), EulerCam using the NGTS filter (fourth), and TRAPPIST 
using an I+z' filter (all remaining light curves). The TRAPPIST data are binned in two-minute intervals.}
\end{figure}
Two additional transit light curves of WASP-49 were obtained using EulerCam at the 1.2~m Euler-Swiss telescope at the La Silla site (Chile).
During the night of 5 December 2012 we observed through a wide (520~nm to 880~nm) filter designed for the upcoming \textit{NGTS} survey \citep{Wheatley13}, 
while during the night of 30 December 2012, an r'-Gunn filter was used. The telescope was slightly defocused for both observations, 
and exposure times were between 35s and 60s (December 5), and 90s (December 30). The data were reduced using relative aperture photometry.
More details on instrument and reduction can be found in \citet{Lendl12}. 

The TRAnsiting Planets and PlanetesImals Small Telescope (TRAPPIST, \citealp{Gillon11a,Jehin11}) is also located at the La Silla site. 
It was used to observe four more transits through an I+z' filter during the nights of 5, 16, and 30 December 2012, and 21 February 2013. 
The exposure times used were 6~s (first two transits) and 10~s (last two transits). The light curves were produced using relative aperture 
photometry, where several apertures were tested and the ideal combination of reference stars was found. IRAF \footnote[1]{IRAF is distributed
by the National Optical Astronomy Observatories, which are operated by the Association of Universities for Research in Astronomy, Inc., under 
cooperative agreement with the National Science Foundation.} was used in the reduction process.

We also included in the analysis the two full transit light curves of each EulerCam and TRAPPIST that have been already described in \cite{Lendl12}.
All broadband light curves are shown in Figure \ref{fig:W49TE}.

\section{Modeling}
\label{sec:W49mod}

\subsection{Method}
To derive the transmission spectrum of the planet and find improved measurements of the planetary and stellar parameters, a Markov chain Monte Carlo 
(MCMC) approach was used. Included in the analysis were all available photometric data as described in Sect. \ref{sec:W49dat} (FORS2, EulerCam, and TRAPPIST). 

We made use of a modified version of the adaptive MCMC code described in detail in \citet{Gillon12a}. In this code, the prescription of 
\citet{Mandel02} is used to model the transit light curves. 
To compensate for correlated noise in the light curves, parametrizations of external variables (such as time, stellar FWHM, and
coordinate shifts) can be included in the photometric baselines models.  These models typically consist of polynomials up to fourth order that are 
multiplied with the theoretical transit light curve. Their coefficients are found by least-squares minimization at every MCMC step. 

\subsubsection{Common noise model}
\label{sec:CNM}
\begin{figure}
 \includegraphics[width=\linewidth]{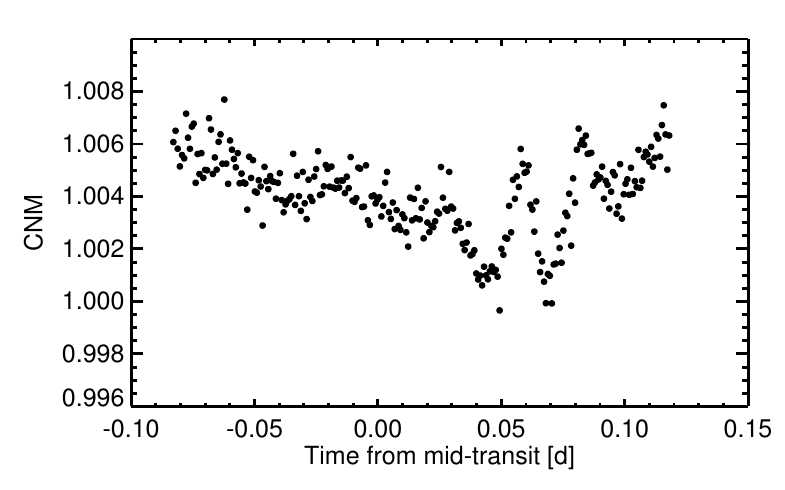}
 \caption{\label{fig:CNM}Example of a CNM, calculated as described in Sect. \ref{sec:CNM}. Here, the CNM of the $\lambda < 788$~nm data of 05 December 2012 is shown.}
 \end{figure}

In this updated version of the code, it is possible to also include a common noise model (CNM) for a set of light curves carrying 
the same correlated noise structure. This CNM is created at each MCMC step by fitting a model transit light curve based on the current 
parameter state together with an invariable, previously determined, normal distribution for the transit depth with a center $dF_{group}$, 
and then co-adding the residuals of this fit. For each time step $t_{i}$, the CNM is calculated as
\begin{equation}
 CNM_{i}=\sum_{k=0}^{n_{lc}} \left( \frac{O_{i,k}}{C_{i,k}}\, w^{-1}_{i,k} \right) \, ,
\end{equation}
where $n_{lc}$ is the total number of light curves the CNM is calculated for, $O_i$ are the observed data, and $C_i$ are the transit model values.
Weights $w_{i,k}$ are attributed according to measurement \mbox{errors $err_{i,k}$,}
\begin{equation}
 w_{i,k}=err^2_{i,k}\, {\sum_{k=0}^{n_{lc}} \frac{1}{err_{i,k}^2}}\, .
\end{equation}
See Fig. \ref{fig:CNM} for an example of a CNM obtained from FORS2 data. This approach is similar to the use of the 
white light curves for the definition of the correlated noise component by \citet{Stevenson14} and others.

\subsubsection{Fitted and fixed parameters}
\label{sec:MCpar}
In the analysis of our combined photometric dataset, the following variables were MCMC fitted (''jump``) parameters:
\begin{itemize}
 \item The impact parameter $b'=a \cos(i_{p})/R_{\ast}$, where $R_{\ast}$ denotes the stellar radius, $a$ the semi-major axis of the planetary orbit, and
$i_{p}$ the orbital inclination.
 \item The transit duration $T_{14}$
 \item The time of mid-transit $T_{0}$
 \item The orbital period $P$
 \item The stellar parameters effective temperature $T_{eff}$ and metallicity [Fe/H].
 \item If desired (as for the test in Sect. \ref{sec:LD}), the linear combination of the quadratic limb-darkening coefficients $(u_1,u_2)$ in each 
       wavelength band, $c_{1,i}=2u_{1,i} + u_{2,i}$ and $c_{2,i} = u_{1,i} - 2u_{2,i}$ \citep{Holman06}.
 \item If a single value of the transit depth is desired (step 1 in our analysis), $dF_{0}=(R_{p}/R_{\ast})^2$ is included as a jump parameter.
 \item If a transmission spectrum is fit (i.e. several values of $dF$, step 2 in our analysis), offsets $ddF_i$ to a pre-defined
       value for $dF_0$.
 \item If a CNM is included, an a priori estimate of the transit depth of each group, $dF_{group}$.
\end{itemize}
The value for the RV amplitude was set to that of the discovery paper.
The eccentricity was set to zero as there has been no evidence for an eccentric orbit of WASP-49b. 
A quadratic model of the form $ I(\mu) = I_{\mathit{center}}\left(1-u_1\left(1-\mu\right)-u_2\left(1-\mu\right)^2\right)$ 
(where $\mu=\cos{\theta}$ and $\theta$ is the angle between the surface normal of the star and the line of sight) was used to account for the effect of stellar limb-darkening
on the transit light curves. The coefficients $u_1$ and $u_2$ were found by interpolating the tables of \citet{Claret11} to match the wavelength bands of our observations. 
The limb-darkening parameters were kept fixed to the values given in Table \ref{tab:LD} throughout our analysis, but we verified our results by allowing for
variable limb-darkening coefficients as described in Sect. \ref{sec:LD}.

\begin{figure*}
  \includegraphics[width=\linewidth]{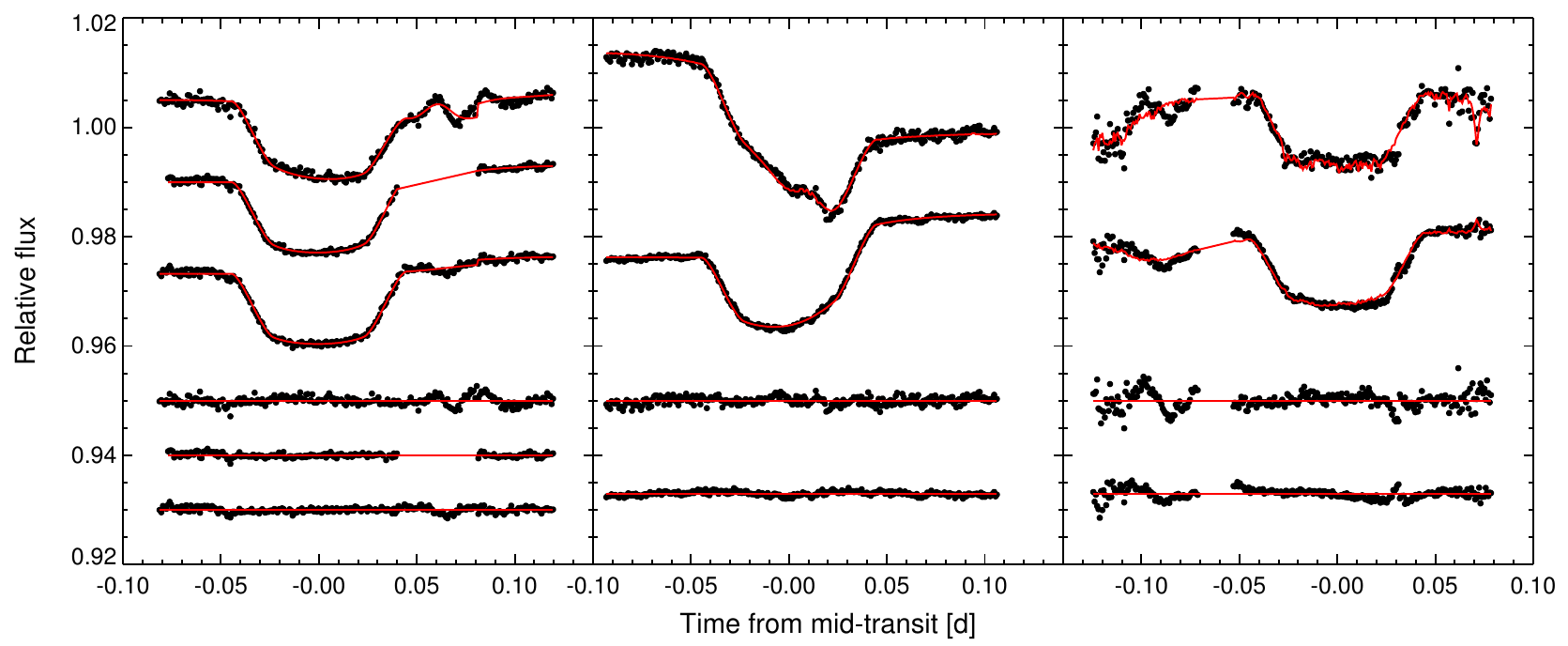}
 \caption{\label{fig:whites} FORS2 spectrophotometry and residuals obtained from using a single bin for each light curve subset, as described in Sect. \ref{sec:step1}. 
 The three subsets of 05 December 2012 are shown in the left panel and correspond to wavelength ranges of (from top to bottom) 738-788, 788-898, and 898-1020 nm.
 The middle and right panels show the two subsets of 14 January 2013 and 07 February 2013, respectively, and correspond to wavelength ranges of (from top to bottom) 738-788,
 and 788-1020 nm. The models are shown as red solid lines and contain the parametrizations of external parameters as described in Table \ref{tab:whites}.
 The residuals are shown below the data which are partially offset for clarity.}
\end{figure*}

\begin{table}
\caption{Limb-darkening coefficients used in the analysis of the photometric data.}   
\label{tab:LD}   
\centering                   
\begin{tabular}{lll}
\hline\hline                 
Wavelength [nm] & $u_1$ & $u_2$ \T \\
\hline       
  738 - 748 & $ 0.318 $ & $ 0.275 $ \T \\
  748 - 758 & $ 0.315 $ & $ 0.273 $  \\
  758 - 768 & $ 0.311 $ & $ 0.272 $  \\
  768 - 778 & $ 0.309 $ & $ 0.27  $ \\
  778 - 788 & $ 0.306 $ & $ 0.268 $  \\
  788 - 798 & $ 0.303 $ & $ 0.267 $  \\
  798 - 808 & $ 0.3   $ & $ 0.265 $  \\
  808 - 818 & $ 0.298 $ & $ 0.264 $  \\
  818 - 828 & $ 0.295 $ & $ 0.262 $  \\
  828 - 838 & $ 0.292 $ & $ 0.261 $  \\
  838 - 848 & $ 0.289 $ & $ 0.26  $ \\
  848 - 858 & $ 0.286 $ & $ 0.259 $  \\
  858 - 868 & $ 0.282 $ & $ 0.258 $  \\
  868 - 878 & $ 0.279 $ & $ 0.258 $  \\
  878 - 888 & $ 0.275 $ & $ 0.257 $  \\
  888 - 898 & $ 0.27  $ & $ 0.258 $  \\
  898 - 908 & $ 0.265 $ & $ 0.258 $  \\
  908 - 918 & $ 0.259 $ & $ 0.259 $  \\
  918 - 928 & $ 0.253 $ & $ 0.26  $ \\
  928 - 938 & $ 0.247 $ & $ 0.261 $  \\
  938 - 948 & $ 0.239 $ & $ 0.263 $  \\
  948 - 958 & $ 0.231 $ & $ 0.266 $  \\
  958 - 968 & $ 0.222 $ & $ 0.269 $  \\
  968 - 978 & $ 0.213 $ & $ 0.272 $  \\
  978 - 988 & $ 0.202 $ & $ 0.276 $  \\
 988 - 1000 & $ 0.189 $ & $ 0.281 $  \\
1000 - 1020 & $ 0.169 $ & $ 0.29  $ \\                
 \textit{r'}  & $  0.28 $ & $ 0.26$ \\      
 \textit{NGTS} & $  0.34$ & $ 0.28$ \\
 \textit{I+z'} & $  0.29$ & $ 0.26$ \\
\hline                       
\end{tabular}
\end{table}

Uniform prior distributions were assumed for most parameters, but for the stellar effective temperature $T_{eff}$ and metallicity [Fe/H], normal
prior distributions were used. These priors were centered on the values of \citet{Lendl12} and their widths were defined as the 1- $\sigma$ errors of these values.
If transmission spectra were derived, a normal prior was applied to $dF_{0}$, with a width of the 1- $\sigma$ errors of the used input value.
We used the method described in \citet{Enoch10} and \citet{Gillon11b} to derive the stellar radius and mass from the transit light curve, stellar 
temperature and metallicity. The contribution (and its error) from the neighboring star was included in the analysis and the transit depths were 
adapted. At each instance, two MCMC chains were run and convergence of the MCMC chains was checked for all results with the Gelman \& Rubin test \citep{Gelman92}. 

\subsubsection{Photometric error adaptation}
The photometric error bars were rescaled by calculating the white $\beta_{w}$ and red $\beta_{r}$ noise scale factors. 
$\beta_{w}$ is given by the ratio of the mean photometric error and the standard deviation of the final 
photometric residuals, and $\beta_{r}$ \citep{Winn08,Gillon10a} is derived by comparing the standard deviations of the binned and unbinned 
residuals. We multiplied our errors with their product $CF = \beta_{r} \times \beta_{w}$ derived from an initial MCMC run of $10^4$ points.
The CF values are given in Table \ref{tab:W49}.

\subsubsection{Baseline models for the FORS2 data}

The FORS2 data are affected by the rotation of the LADC with respect to the detector and therefore show a strong noise 
component correlated with the parallactic angle. For these light curves, we tested baseline models involving the parallactic angle 
$\mathit{par}$, the change in parallactic angle $\mathit{dpar}$ (i.e., the derotator speed), and the trigonometric functions $\sin(\mathit{par})$ and $\cos(\mathit{par})$.
Including the parallactic angle in baseline models for the FORS2 data leads to a much better fit of the overall light curve shapes than time-dependent
models. Similarly, including the CNM in the photometric baseline produces very good fits to the data, efficiently accounting for short-term
photometric variations that consistently appear in sets of light curves, but cannot be modeled as simple dependencies on external parameters.

\subsection{Step-by-step procedure}

We carried out the analysis with the aim to first infer the most reliable analysis method for the FORS2 data, and then applied it to obtain
an accurate transmission spectrum for WASP-49b. To do so, we proceeded in the steps outlined below.

\subsubsection{Step 1: An overall transit depth}
\label{sec:step1}
We first identified the light curves that are affected by the same photometric variations and hence should form the subsets possessing the
same CNM. For all three dates, a clear division is seen between the light curves obtained with two reference stars (the five shortest-wavelength bins, as well
as the light curve centered on the K feature), and all other light curves. This is a clear consequence of the LADC being at the root of the strongest
correlated noise source, as spatially varying transmission affects each reference star differently. For the data of 05 December 2012, we further subdivided the
remaining light curves into two groups, light curves between 788 and 898~nm, which have fewer points because of detector saturation, and 
the light curves at wavelengths above 898~nm, where counts remained in the linear detector range.

For each subset, we combined all data to produce a ``white'' light curve. We then tested a number of photometric baseline models for these white 
light curves, finding the best modeling of these curves by accounting for large trends by means of second-order polynomials in $\mathit{par}$ and 
modeling the short-timescale variations in the $\lambda<788$~nm light curves using fourth-order polynomials in $\cos(par)$ or $\mathit{dpar}$, or 
a first-order polynomial in the stellar FWHM. These baseline functions are listed in Table \ref{tab:whites}. For the 05 December 2012 data, 
an offset is included at the change of exposure time. We then performed a combined MCMC analysis on all white light curves, allowing for a 
single transit depth. The result obtained is our best absolute transit depth $dF_{0}$, which was then used to calculate the 
transmission spectrum. These light curves are shown in Fig. \ref{fig:whites}. We find a resulting value for the transit depth of $dF_0=0.0133 \pm 0.0002$.

\begin{table}
 \caption{\label{tab:whites}Baseline functions used in step 1 of our analysis, where a single transit depth is inferred from the binned data of
 all light curve subsets. The baseline functions of the form $p^j(i)$ denote a polynomial of order $j$ in parameter $i$, where $i$ can be
 the parallactic angle $par$, its cosine $\cos(par)$, the differential parallactic angle from one exposure to the next $\mathit{dpar}$, 
 and the PSF or spectral full-width at half maximum $\mathit{fwhm}$. $\mathit{off}$ refers to an offset included at the change in exposure time
 for the 05 December 2012 observations.}
\begin{tabular}{lll}
\hline\hline 
Wavelength [nm] & Date & Baseline function  \T \\  
\hline
738 - 788 & 05 Dec 2012 & $p^4(\cos(\mathit{par}))+\mathit{off}$ \T \\
788 - 898 & 05 Dec 2012 & $p^2(\mathit{par})+\mathit{off}$ \\
898 - 1020 & 05 Dec 2012 & $p^2(\mathit{par})+\mathit{off}$ \\
738 - 788 & 14 Jan 2013 & $p^2(\mathit{par})+p^4(\mathit{dpar})$ \\
738 - 1020 & 14 Jan 2013 & $p^2(\mathit{par})$ \\
738 - 788 & 07 Feb 2013 & $p^1(\mathit{fwhm})+p^2(\mathit{par})$ \\
738 - 1020 & 07 Feb 2012 & $p^1(\mathit{fwhm})+p^2(\mathit{par})$ \\
\hline
\end{tabular}
\end{table}

\subsubsection{Step 2: Individual transmission spectra}
\begin{figure*}
 \includegraphics[width=\linewidth]{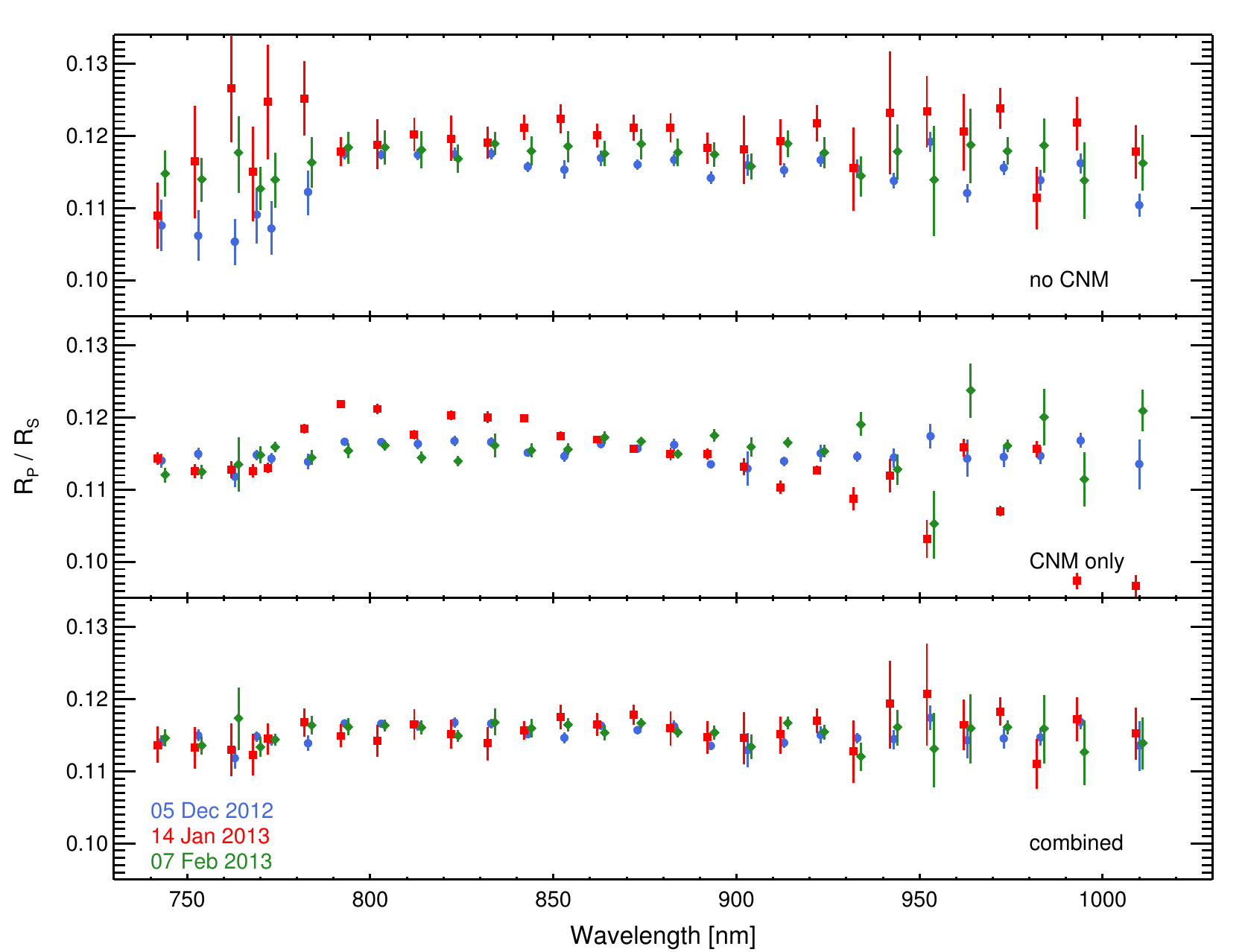}
 \caption{\label{fig:ssep}FORS2 transmission spectra of the three dates obtained with three baseline modeling approaches: no CNM, but parametrizations of external parameters 
 as given in Sect. \ref{sec:step2} (top), CNM alone (middle), and CNM together with basic parametrizations as detailed in Sect. \ref{sec:step2} (bottom). 
 Blue filled circles denote the FORS2 transmission spectrum of 05 December 2012, red squares that of 14 January 2013, and green diamonds that of 07 February 2013.
 In this plot, the spectra of 14 January 2013 and 07 February 2013 are offset from their nominal wavelengths by 1~nm to avoid overlaps.}
\end{figure*}
\label{sec:step2}
We then derived individual transmission spectra for each date. Keeping an a priory transit depth fixed to the previously 
obtained $dF_0$, we inferred offsets $ddF_i$ from this value for each light curve. At this point, we searched for the best baseline models 
for each previously defined light curve set, testing three approaches, that relied to various degrees on the CNM.

\begin{enumerate}
 \item \textbf{No CNM:} the photometric baselines consist solely of functions of external parameters, that is, $par$, $\cos(\mathit{par})$, $\mathit{dpar}$, and $\mathit{FWHM}$.
   These functions, and the wavelength range for light curves to which they are applied, are the same as inferred in Sect. \ref{sec:step1} and listed 
   in Table \ref{tab:whites}.
 \item \textbf{CNM only:} no parametrizations of external variables are used, except for the offset at the exposure time change for the 05 December 2012 data. 
 All light curves are fit with a first-order polynomial of the CNM (i.e., a function of the form $a_0+a_1\,\mathit{CNM}$) only. As mentioned in Sect. \ref{sec:CNM}, 
 the CNMs are calculated based on an input distribution 
 for the transit depth. We used a Gaussian centered on the previously obtained $dF_0$, with a width of the 1-$\sigma$ error bars on $dF_0$. 
 Higher-order polynomials with respect to the CNM were tested but did not improve the results.
  \item \textbf{Combined:} the photometric baseline functions include the CNM and low-order polynomials of external parameters. 
   The best baseline models were found to consist of first-order polynomials of the CNM, together with second-order
   polynomials of $\mathit{par}$ (January and February data).
 \end{enumerate}
The resulting transmission spectra are shown in Fig. \ref{fig:ssep}: the results obtained without CNM (top panel) show larger error bars than the other 
modeling approaches, while the data are noisier at low wavelengths and show overall offsets between the three dates, most remarkably a median offset of 0.005
in $R_p/R_{\ast}$ between the data of 06 December 2012 and 14 January 2013. The transmission spectra inferred from CNM-only models (middle panel) have greatly 
reduced error bars, but the spectra inferred from the three dates do not agree, with substantial scatter at 
long wavelengths. The 14 January 2013 data show a distinct slope between 790 and 900~nm, a structure not reproduced for the other dates. 
The best agreement between the spectra from the three dates is found if the CNM and low-order parametrizations of external parameters are used together (third
panel), and we used this approach to derive of our final transmission spectrum.

\subsubsection{Step 3: A combined transmission spectrum}
\label{sec:st3}
The final transmission spectrum was inferred from a global analysis of all available photometric data: the three FORS2 observations, and all available 
additional EulerCam and TRAPPIST broadband photometry. As the latter are light curves obtained with different instruments at different dates, CNMs cannot be
applied to these data. The light curves are displayed in Fig. \ref{fig:W49TE} (broadband), and Fig. \ref{fig:W49LC}, 
and the respective baseline functions are given in Table \ref{tab:W49}. For the analysis of the FORS2 data, we chose the third option 
outlined above: using CNMs together with functions of $par$ (and an offset at the change of exposure time in the 5 December 2012 data). 
The light curve subsets contributing to each CNM are the seven subsets defined above 
(three for the 5 December 2012 data, and two each for the 14 January 2013 and 7 February 2013 data). The analysis consisted of 
two MCMC chains of $10^5$ points each, allowing only for unique values of the transit depths for each wavelength bin. The resulting transmission spectrum is 
shown in Fig. \ref{fig:mod}.

\subsubsection{Stellar limb-darkening}
\label{sec:LD}
As stellar limb-darkening affects the transit shape, we decided to verify our transmission spectrum against variation in the limb-darkening coefficients. To do so,
we carried out additional global analyses (identical to those described in Sect. \ref{sec:st3}), while allowing the limb-darkening coefficients to vary, assuming
a normal prior distribution for them. This prior was centered on the interpolated value and had a width large enough to encompass the values of the
neighboring passbands at 1-$\sigma$. The results are consistent with our previously derived transmission spectrum: when a prior is included, the maximum
offset between the two runs is $0.14\sigma$ and the transmission spectra uncertainties are similar. 
\begin{longtab}
\begin{longtable}{lllllll}
\caption{\label{tab:W49}Details on the observations: wavelength band, date, baseline parameters and noise statistics of all data included in the global 
analysis of WASP-49b. The baseline functions of the form $p^j(i)$ denote a polynomial of order $j$ in parameter $i$, where $i$ can be
time $t$, parallactic angle $\mathit{par}$, the sky background $\mathit{sky}$ and the PSF or spectral full-width at half maximum $\mathit{fwhm}$. 
$\mathit{off}$ refers to an offset due to the change
in exposure time on FORS2, or a telescope meridian flip for some TRAPPIST light curves. The red- and white noise amplitudes $\beta_r$ and $\beta_w$, 
the error adaptation factor $\mathit{CF}$, and the RMS is given for data binned in bins of two minutes. For the FORS2 data, the four data quality parameters are given 
for the global fit (left value) and for fits restricted to single transit events (right value).}  \\             
\hline\hline 
Wavelength [nm] & Date      & Baseline function  &  \qquad $\beta_r$  &  \qquad$\beta_w$    &   \qquad CF         &  $\mathrm{RMS_{120s}}$ [\%] \T \\  
\hline
  FORS2      &              &   &             &              &              &              \T \\
   738 - 748 &  05 Dec 2012 & $p^1(\mathit{CNM})+\mathit{off}               $  & $1.33$ \quad $1.33$ & $0.97$ \quad $0.97$ & $1.29$ \quad $1.29$ & $0.080$ \quad $0.081$ \\
             &  14 Jan 2013 & $p^1(\mathit{CNM})+p^2(\mathit{par})          $  & $1.12$ \quad $1.15$ & $1.00$ \quad $0.99$ & $1.12$ \quad $1.14$ & $0.074$ \quad $0.087$ \\
             &  07 Feb 2013 & $p^1(\mathit{CNM})+p^2(\mathit{par})          $  & $1.98$ \quad $1.18$ & $0.88$ \quad $0.90$ & $1.74$ \quad $1.06$ & $0.081$ \quad $0.093$ \\
   748 - 758 &  05 Dec 2012 & $p^1(\mathit{CNM})+\mathit{off}               $  & $1.24$ \quad $1.13$ & $0.81$ \quad $0.99$ & $1.01$ \quad $1.12$ & $0.055$ \quad $0.074$ \\
             &  14 Jan 2013 & $p^1(\mathit{CNM})+p^2(\mathit{par})          $  & $1.58$ \quad $1.27$ & $1.11$ \quad $0.97$ & $1.75$ \quad $1.23$ & $0.083$ \quad $0.088$ \\
             &  07 Feb 2013 & $p^1(\mathit{CNM})+p^2(\mathit{par})          $  & $1.50$ \quad $1.20$ & $0.52$ \quad $0.87$ & $0.77$ \quad $1.04$ & $0.022$ \quad $0.083$ \\
   758 - 768 &  05 Dec 2012 & $p^1(\mathit{CNM})+\mathit{off}               $  & $1.39$ \quad $2.01$ & $0.70$ \quad $0.86$ & $0.98$ \quad $1.74$ & $0.044$ \quad $0.082$ \\
             &  14 Jan 2013 & $p^1(\mathit{CNM})+p^2(\mathit{par})          $  & $1.02$ \quad $1.50$ & $0.71$ \quad $0.92$ & $0.73$ \quad $1.37$ & $0.045$ \quad $0.093$ \\
             &  07 Feb 2013 & $p^1(\mathit{CNM})+p^2(\mathit{par})          $  & $1.03$ \quad $3.71$ & $0.71$ \quad $0.91$ & $0.73$ \quad $3.36$ & $0.045$ \quad $0.122$ \\
   768 - 778 &  05 Dec 2012 & $p^1(\mathit{CNM})+\mathit{off}               $  & $1.48$ \quad $1.25$ & $0.69$ \quad $0.81$ & $1.02$ \quad $1.01$ & $0.044$ \quad $0.054$ \\
             &  14 Jan 2013 & $p^1(\mathit{CNM})+p^2(\mathit{par})          $  & $1.49$ \quad $1.18$ & $0.66$ \quad $0.82$ & $0.98$ \quad $0.97$ & $0.043$ \quad $0.064$ \\
             &  07 Feb 2013 & $p^1(\mathit{CNM})+p^2(\mathit{par})          $  & $1.31$ \quad $1.00$ & $0.72$ \quad $0.76$ & $0.95$ \quad $0.76$ & $0.045$ \quad $0.058$ \\
   778 - 788 &  05 Dec 2012 & $p^1(\mathit{CNM})+\mathit{off}               $  & $1.00$ \quad $1.57$ & $0.68$ \quad $1.11$ & $0.68$ \quad $1.75$ & $0.042$ \quad $0.081$ \\
             &  14 Jan 2013 & $p^1(\mathit{CNM})+p^2(\mathit{par})          $  & $1.41$ \quad $1.09$ & $0.72$ \quad $0.93$ & $1.02$ \quad $1.01$ & $0.052$ \quad $0.070$ \\
             &  07 Feb 2013 & $p^1(\mathit{CNM})+p^2(\mathit{par})          $  & $0.97$ \quad $1.49$ & $0.77$ \quad $0.87$ & $0.80$ \quad $1.29$ & $0.047$ \quad $0.078$ \\
 \textit{765 - 773 (K)} & 05 Dec 2012  &  $p^1(\mathit{CNM})+\mathit{off}       $  & $3.68$ \quad $1.42$ & $1.07$ \quad $0.69$ & $3.94$ \quad $0.98$ & $0.077$ \quad $0.044$ \\
             &  14 Jan 2013 & $p^1(\mathit{CNM})+p^2(\mathit{par})          $  & $1.39$ \quad $1.50$ & $0.81$ \quad $0.76$ & $1.13$ \quad $1.15$ & $0.047$ \quad $0.068$ \\
             &  07 Feb 2013 & $p^1(\mathit{CNM})+p^2(\mathit{par})          $  & $2.51$ \quad $1.57$ & $0.80$ \quad $0.73$ & $2.01$ \quad $1.14$ & $0.056$ \quad $0.078$ \\
   788 - 798 &  05 Dec 2012 & $p^1(\mathit{CNM})+\mathit{off}               $  & $1.09$ \quad $1.02$ & $1.03$ \quad $0.72$ & $1.12$ \quad $0.73$ & $0.063$ \quad $0.046$ \\
             &  14 Jan 2013 & $p^1(\mathit{CNM})+p^2(\mathit{par})          $  & $1.87$ \quad $1.10$ & $1.05$ \quad $0.82$ & $1.96$ \quad $0.91$ & $0.085$ \quad $0.051$ \\
             &  07 Feb 2013 & $p^1(\mathit{CNM})+p^2(\mathit{par})          $  & $2.50$ \quad $1.71$ & $1.01$ \quad $0.78$ & $2.53$ \quad $1.34$ & $0.085$ \quad $0.064$ \\
   798 - 808 &  05 Dec 2012 & $p^1(\mathit{CNM})+\mathit{off}               $  & $3.50$ \quad $1.04$ & $1.03$ \quad $0.70$ & $3.61$ \quad $0.73$ & $0.097$ \quad $0.045$ \\
             &  14 Jan 2013 & $p^1(\mathit{CNM})+p^2(\mathit{par})          $  & $2.00$ \quad $1.69$ & $0.96$ \quad $0.76$ & $1.93$ \quad $1.29$ & $0.078$ \quad $0.047$ \\
             &  07 Feb 2013 & $p^1(\mathit{CNM})+p^2(\mathit{par})          $  & $1.53$ \quad $1.19$ & $0.99$ \quad $0.73$ & $1.52$ \quad $0.87$ & $0.080$ \quad $0.056$ \\
   808 - 818 &  05 Dec 2012 & $p^1(\mathit{CNM})+\mathit{off}               $  & $1.39$ \quad $1.48$ & $1.04$ \quad $0.69$ & $1.45$ \quad $1.02$ & $0.080$ \quad $0.044$ \\
             &  14 Jan 2013 & $p^1(\mathit{CNM})+p^2(\mathit{par})          $  & $3.72$ \quad $1.54$ & $1.20$ \quad $0.76$ & $4.46$ \quad $1.18$ & $0.116$ \quad $0.048$ \\
             &  07 Feb 2013 & $p^1(\mathit{CNM})+p^2(\mathit{par})          $  & $1.14$ \quad $1.45$ & $1.00$ \quad $0.74$ & $1.14$ \quad $1.08$ & $0.088$ \quad $0.061$ \\
   818 - 828 &  05 Dec 2012 & $p^1(\mathit{CNM})+\mathit{off}               $  & $1.27$ \quad $1.47$ & $0.97$ \quad $0.66$ & $1.23$ \quad $0.98$ & $0.088$ \quad $0.043$ \\
             &  14 Jan 2013 & $p^1(\mathit{CNM})+p^2(\mathit{par})          $  & $1.49$ \quad $1.48$ & $0.92$ \quad $0.79$ & $1.37$ \quad $1.17$ & $0.092$ \quad $0.053$ \\
             &  07 Feb 2013 & $p^1(\mathit{CNM})+p^2(\mathit{par})          $  & $1.18$ \quad $1.27$ & $0.82$ \quad $0.70$ & $0.97$ \quad $0.90$ & $0.064$ \quad $0.051$ \\
   828 - 838 &  05 Dec 2012 & $p^1(\mathit{CNM})+\mathit{off}               $  & $1.09$ \quad $1.29$ & $0.93$ \quad $0.73$ & $1.01$ \quad $0.94$ & $0.070$ \quad $0.046$ \\
             &  14 Jan 2013 & $p^1(\mathit{CNM})+p^2(\mathit{par})          $  & $1.47$ \quad $1.63$ & $0.66$ \quad $0.85$ & $0.97$ \quad $1.38$ & $0.055$ \quad $0.053$ \\
             &  07 Feb 2013 & $p^1(\mathit{CNM})+p^2(\mathit{par})          $  & $1.50$ \quad $2.75$ & $0.77$ \quad $0.78$ & $1.15$ \quad $2.13$ & $0.068$ \quad $0.065$ \\
   838 - 848 &  05 Dec 2012 & $p^1(\mathit{CNM})+\mathit{off}               $  & $1.11$ \quad $1.00$ & $0.82$ \quad $0.68$ & $0.91$ \quad $0.68$ & $0.052$ \quad $0.042$ \\
             &  14 Jan 2013 & $p^1(\mathit{CNM})+p^2(\mathit{par})          $  & $1.69$ \quad $1.05$ & $0.77$ \quad $0.72$ & $1.29$ \quad $0.76$ & $0.048$ \quad $0.038$ \\
             &  07 Feb 2013 & $p^1(\mathit{CNM})+p^2(\mathit{par})          $  & $1.56$ \quad $1.90$ & $0.76$ \quad $0.73$ & $1.18$ \quad $1.38$ & $0.048$ \quad $0.058$ \\
   848 - 858 &  05 Dec 2012 & $p^1(\mathit{CNM})+\mathit{off}               $  & $1.48$ \quad $1.43$ & $0.79$ \quad $0.71$ & $1.17$ \quad $1.02$ & $0.053$ \quad $0.052$ \\
             &  14 Jan 2013 & $p^1(\mathit{CNM})+p^2(\mathit{par})          $  & $1.63$ \quad $1.29$ & $0.85$ \quad $0.78$ & $1.38$ \quad $1.00$ & $0.054$ \quad $0.047$ \\
             &  07 Feb 2013 & $p^1(\mathit{CNM})+p^2(\mathit{par})          $  & $1.05$ \quad $1.57$ & $0.72$ \quad $0.68$ & $0.76$ \quad $1.07$ & $0.038$ \quad $0.051$ \\
   858 - 868 &  05 Dec 2012 & $p^1(\mathit{CNM})+\mathit{off}               $  & $1.28$ \quad $1.06$ & $0.78$ \quad $0.76$ & $1.00$ \quad $0.80$ & $0.047$ \quad $0.046$ \\
             &  14 Jan 2013 & $p^1(\mathit{CNM})+p^2(\mathit{par})          $  & $1.26$ \quad $1.25$ & $0.74$ \quad $0.74$ & $0.93$ \quad $0.93$ & $0.046$ \quad $0.046$ \\
             &  07 Feb 2013 & $p^1(\mathit{CNM})+p^2(\mathit{par})          $  & $1.06$ \quad $1.83$ & $0.77$ \quad $0.64$ & $0.81$ \quad $1.17$ & $0.045$ \quad $0.051$ \\
   868 - 878 &  05 Dec 2012 & $p^1(\mathit{CNM})+\mathit{off}               $  & $1.80$ \quad $1.05$ & $0.80$ \quad $0.72$ & $1.44$ \quad $0.75$ & $0.048$ \quad $0.047$ \\
             &  14 Jan 2013 & $p^1(\mathit{CNM})+p^2(\mathit{par})          $  & $1.54$ \quad $1.06$ & $0.84$ \quad $0.77$ & $1.29$ \quad $0.81$ & $0.056$ \quad $0.045$ \\
             &  07 Feb 2013 & $p^1(\mathit{CNM})+p^2(\mathit{par})          $  & $2.02$ \quad $1.22$ & $0.98$ \quad $0.66$ & $1.98$ \quad $0.81$ & $0.079$ \quad $0.045$ \\
   878 - 888 &  05 Dec 2012 & $p^1(\mathit{CNM})+\mathit{off}               $  & $1.73$ \quad $1.56$ & $0.84$ \quad $0.76$ & $1.45$ \quad $1.19$ & $0.063$ \quad $0.047$ \\
             &  14 Jan 2013 & $p^1(\mathit{CNM})+p^2(\mathit{par})          $  & $1.17$ \quad $1.80$ & $0.80$ \quad $0.80$ & $0.94$ \quad $1.44$ & $0.054$ \quad $0.048$ \\
             &  07 Feb 2013 & $p^1(\mathit{CNM})+p^2(\mathit{par})          $  & $1.89$ \quad $1.11$ & $1.13$ \quad $0.69$ & $2.14$ \quad $0.77$ & $0.094$ \quad $0.052$ \\
   888 - 898 &  05 Dec 2012 & $p^1(\mathit{CNM})+\mathit{off}               $  & $2.57$ \quad $1.00$ & $1.21$ \quad $0.75$ & $3.11$ \quad $0.75$ & $0.105$ \quad $0.048$ \\
             &  14 Jan 2013 & $p^1(\mathit{CNM})+p^2(\mathit{par})          $  & $2.79$ \quad $1.53$ & $1.20$ \quad $0.84$ & $3.35$ \quad $1.29$ & $0.103$ \quad $0.056$ \\
             &  07 Feb 2013 & $p^1(\mathit{CNM})+p^2(\mathit{par})          $  & $1.67$ \quad $1.48$ & $0.94$ \quad $0.77$ & $1.57$ \quad $1.13$ & $0.084$ \quad $0.063$ \\
   898 - 908 &  05 Dec 2012 & $p^1(\mathit{CNM})+\mathit{off}               $  & $1.00$ \quad $3.66$ & $0.94$ \quad $1.08$ & $0.94$ \quad $3.94$ & $0.080$ \quad $0.077$ \\
             &  14 Jan 2013 & $p^1(\mathit{CNM})+p^2(\mathit{par})          $  & $1.45$ \quad $2.03$ & $1.05$ \quad $0.98$ & $1.52$ \quad $1.98$ & $0.086$ \quad $0.079$ \\
             &  07 Feb 2013 & $p^1(\mathit{CNM})+p^2(\mathit{par})          $  & $1.24$ \quad $2.12$ & $1.08$ \quad $0.81$ & $1.34$ \quad $1.71$ & $0.094$ \quad $0.077$ \\
   908 - 918 &  05 Dec 2012 & $p^1(\mathit{CNM})+\mathit{off}               $  & $1.30$ \quad $1.39$ & $1.21$ \quad $0.82$ & $1.57$ \quad $1.13$ & $0.109$ \quad $0.046$ \\
             &  14 Jan 2013 & $p^1(\mathit{CNM})+p^2(\mathit{par})          $  & $1.17$ \quad $1.73$ & $0.90$ \quad $0.84$ & $1.06$ \quad $1.45$ & $0.094$ \quad $0.063$ \\
             &  07 Feb 2013 & $p^1(\mathit{CNM})+p^2(\mathit{par})          $  & $1.20$ \quad $1.30$ & $0.87$ \quad $0.68$ & $1.04$ \quad $0.88$ & $0.083$ \quad $0.051$ \\
   918 - 928 &  05 Dec 2012 & $p^1(\mathit{CNM})+\mathit{off}               $  & $3.56$ \quad $2.49$ & $0.94$ \quad $0.81$ & $3.36$ \quad $2.01$ & $0.139$ \quad $0.057$ \\
             &  14 Jan 2013 & $p^1(\mathit{CNM})+p^2(\mathit{par})          $  & $1.00$ \quad $1.17$ & $0.76$ \quad $0.81$ & $0.76$ \quad $0.94$ & $0.058$ \quad $0.054$ \\
             &  07 Feb 2013 & $p^1(\mathit{CNM})+p^2(\mathit{par})          $  & $1.47$ \quad $1.36$ & $0.88$ \quad $0.78$ & $1.29$ \quad $1.07$ & $0.078$ \quad $0.074$ \\
   928 - 938 &  05 Dec 2012 & $p^1(\mathit{CNM})+\mathit{off}               $  & $2.50$ \quad $1.08$ & $0.62$ \quad $1.03$ & $1.55$ \quad $1.12$ & $0.068$ \quad $0.063$ \\
             &  14 Jan 2013 & $p^1(\mathit{CNM})+p^2(\mathit{par})          $  & $1.56$ \quad $1.91$ & $0.73$ \quad $1.12$ & $1.14$ \quad $2.14$ & $0.079$ \quad $0.093$ \\
             &  07 Feb 2013 & $p^1(\mathit{CNM})+p^2(\mathit{par})          $  & $1.73$ \quad $1.62$ & $0.78$ \quad $1.13$ & $1.34$ \quad $1.83$ & $0.063$ \quad $0.116$ \\
   938 - 948 &  05 Dec 2012 & $p^1(\mathit{CNM})+\mathit{off}               $  & $1.18$ \quad $1.89$ & $0.74$ \quad $1.04$ & $0.87$ \quad $1.96$ & $0.057$ \quad $0.085$ \\
             &  14 Jan 2013 & $p^1(\mathit{CNM})+p^2(\mathit{par})          $  & $1.46$ \quad $2.57$ & $0.74$ \quad $1.21$ & $1.08$ \quad $3.11$ & $0.060$ \quad $0.105$ \\
             &  07 Feb 2013 & $p^1(\mathit{CNM})+p^2(\mathit{par})          $  & $1.28$ \quad $2.02$ & $0.70$ \quad $1.13$ & $0.90$ \quad $2.27$ & $0.051$ \quad $0.120$ \\
   948 - 958 &  05 Dec 2012 & $p^1(\mathit{CNM})+\mathit{off}               $  & $2.76$ \quad $2.51$ & $0.77$ \quad $1.01$ & $2.13$ \quad $2.53$ & $0.066$ \quad $0.085$ \\
             &  14 Jan 2013 & $p^1(\mathit{CNM})+p^2(\mathit{par})          $  & $1.88$ \quad $2.79$ & $0.73$ \quad $1.20$ & $1.38$ \quad $3.35$ & $0.058$ \quad $0.103$ \\
             &  07 Feb 2013 & $p^1(\mathit{CNM})+p^2(\mathit{par})          $  & $1.56$ \quad $3.67$ & $0.69$ \quad $1.29$ & $1.07$ \quad $4.73$ & $0.052$ \quad $0.158$ \\
   958 - 968 &  05 Dec 2012 & $p^1(\mathit{CNM})+\mathit{off}               $  & $1.83$ \quad $3.54$ & $0.64$ \quad $1.02$ & $1.17$ \quad $3.61$ & $0.049$ \quad $0.094$ \\
             &  14 Jan 2013 & $p^1(\mathit{CNM})+p^2(\mathit{par})          $  & $1.22$ \quad $1.67$ & $0.66$ \quad $0.94$ & $0.81$ \quad $1.57$ & $0.047$ \quad $0.084$ \\
             &  07 Feb 2013 & $p^1(\mathit{CNM})+p^2(\mathit{par})          $  & $1.12$ \quad $3.73$ & $0.69$ \quad $1.14$ & $0.77$ \quad $4.24$ & $0.052$ \quad $0.155$ \\
   968 - 978 &  05 Dec 2012 & $p^1(\mathit{CNM})+\mathit{off}               $  & $1.46$ \quad $1.99$ & $0.77$ \quad $0.97$ & $1.13$ \quad $1.93$ & $0.063$ \quad $0.077$ \\
             &  14 Jan 2013 & $p^1(\mathit{CNM})+p^2(\mathit{par})          $  & $2.11$ \quad $1.00$ & $0.81$ \quad $0.94$ & $1.71$ \quad $0.94$ & $0.076$ \quad $0.080$ \\
             &  07 Feb 2013 & $p^1(\mathit{CNM})+p^2(\mathit{par})          $  & $1.21$ \quad $1.00$ & $0.71$ \quad $0.88$ & $0.86$ \quad $0.88$ & $0.054$ \quad $0.095$ \\
   978 - 988 &  05 Dec 2012 & $p^1(\mathit{CNM})+\mathit{off}               $  & $1.37$ \quad $1.54$ & $0.78$ \quad $0.99$ & $1.07$ \quad $1.52$ & $0.073$ \quad $0.079$ \\
             &  14 Jan 2013 & $p^1(\mathit{CNM})+p^2(\mathit{par})          $  & $1.62$ \quad $1.46$ & $1.13$ \quad $1.04$ & $1.83$ \quad $1.52$ & $0.115$ \quad $0.085$ \\
             &  07 Feb 2013 & $p^1(\mathit{CNM})+p^2(\mathit{par})          $  & $2.01$ \quad $3.49$ & $1.13$ \quad $1.21$ & $2.27$ \quad $4.24$ & $0.119$ \quad $0.157$ \\
  988 - 1000 &  05 Dec 2012 & $p^1(\mathit{CNM})+\mathit{off}               $  & $3.61$ \quad $1.40$ & $1.31$ \quad $1.04$ & $4.73$ \quad $1.45$ & $0.158$ \quad $0.080$ \\
             &  14 Jan 2013 & $p^1(\mathit{CNM})+p^2(\mathit{par})          $  & $3.66$ \quad $1.24$ & $1.16$ \quad $1.08$ & $4.24$ \quad $1.34$ & $0.147$ \quad $0.094$ \\
             &  07 Feb 2013 & $p^1(\mathit{CNM})+p^2(\mathit{par})          $  & $1.00$ \quad $3.07$ & $0.88$ \quad $1.28$ & $0.88$ \quad $3.91$ & $0.094$ \quad $0.150$ \\
 1000 - 1020 &  05 Dec 2012 & $p^1(\mathit{CNM})+\mathit{off}               $  & $3.48$ \quad $3.75$ & $1.22$ \quad $1.19$ & $4.24$ \quad $4.46$ & $0.150$ \quad $0.115$ \\
             &  14 Jan 2013 & $p^1(\mathit{CNM})+p^2(\mathit{par})          $  & $3.05$ \quad $1.30$ & $1.28$ \quad $1.21$ & $3.91$ \quad $1.57$ & $0.153$ \quad $0.110$ \\
             &  07 Feb 2013 & $p^1(\mathit{CNM})+p^2(\mathit{par})          $  & $2.84$ \quad $2.81$ & $1.12$ \quad $1.13$ & $3.18$ \quad $3.18$ & $0.134$ \quad $0.135$ \\
\hline                                                                                                                                             
EulerCam     &              &                    &          &             &             &         \T \\                                            
  r'-Gunn    &  19 Mar 2011 & $p^2(t)+p(sky^1)          $ & $1.45$ \quad - & $1.25$ \quad  -  & $1.81$ \quad  -  & $0.066$ \quad -  \\                  
  r'-Gunn    &  24 Mar 2011 & $p^2(t)                   $ & $1.61$ \quad - & $1.32$ \quad  -  & $2.13$ \quad  -  & $0.090$ \quad -  \\                  
  r'-Gunn    &  30 Dec 2012 & $p^2(t)                   $ & $1.95$ \quad - & $1.68$ \quad  -  & $3.28$ \quad  -  & $0.089$ \quad -  \\                  
  NGTS       &  05 Dec 2012 & $p^2(t)+p^1(\mathit{fwhm})$ & $1.27$ \quad - & $1.32$ \quad  -  & $1.68$ \quad  -  & $0.066$ \quad -  \\                  
\hline                                                                                                                                             
TRAPPIST     &              &                 &            &             &              &         \T \\                                            
  I+z'       &  19 Jan 2011 & $p^2(t)                   $  & $1.15$ \quad  -  & $1.12$ \quad  -  & $1.29$ \quad  -  & $0.150$ \quad -  \\                
  I+z'       &  24 Oct 2011 & $p^2(t)                   $  & $1.80$ \quad  -  & $1.16$ \quad  -  & $2.09$ \quad  -  & $0.158$ \quad -  \\                
  I+z'       &  05 Dec 2012 & $p^2(t) + \mathit{off}    $  & $1.12$ \quad  -  & $1.04$ \quad  -  & $1.17$ \quad  -  & $0.141$ \quad -  \\
  I+z'       &  16 Dec 2012 & $p^2(t)                   $  & $2.07$ \quad  -  & $1.11$ \quad  -  & $2.30$ \quad  -  & $0.163$ \quad -  \\
  I+z'       &  30 Dec 2012 & $p^2(t) + \mathit{off}    $  & $1.18$ \quad  -  & $1.16$ \quad  -  & $1.37$ \quad  -  & $0.165$ \quad -  \\
  I+z'       &  21 Feb 2013 & $p^2(t) + \mathit{off}    $  & $1.27$ \quad  -  & $0.91$ \quad  -  & $1.15$ \quad  -  & $0.111$ \quad -  \\
\hline
\end{longtable}
\end{longtab}

\section{Results}
\label{sec:res}
\subsection{Individual transits}
From the analysis of three sets of FORS2 data, we obtained a set of independently derived transmission spectra of Wasp-49b between 730 and 1020~nm. 

\begin{figure}
 \includegraphics[width=\linewidth]{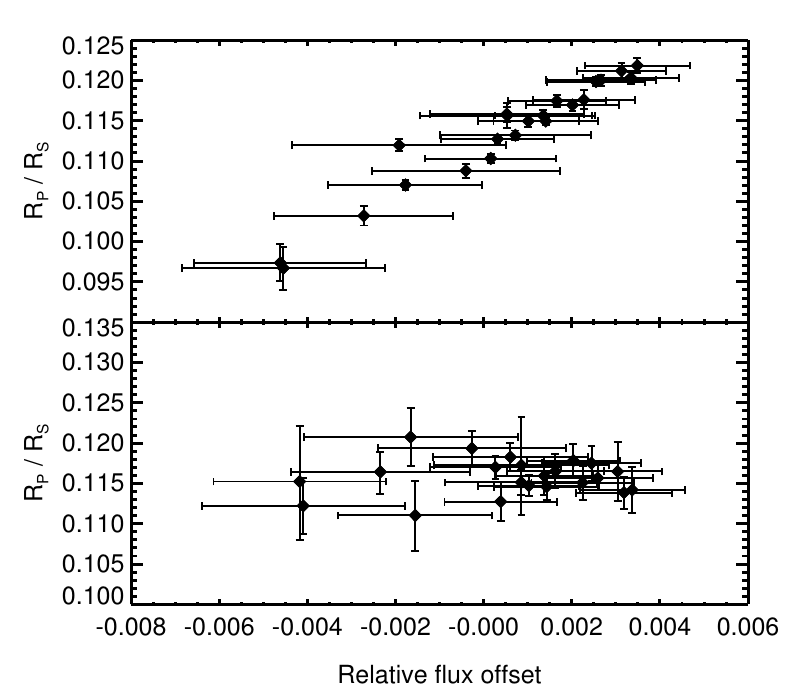}
 \caption{\label{fig:ofc}$R_p/R_\ast$ values inferred from the CNM-only analysis (top) and the combined analysis (bottom) of the $\lambda>788$~nm 
 light curves of 14 January 2013 against the trend amplitude (pre-transit - post-transit flux). 
 A highly significant correlation ($p=0.97)$ is easily visible in the upper panel but is lacking in the lower panel.
 }
\end{figure}

To evaluate the reliability of the derived spectra, we tested three approaches for modeling systematic noise: the exclusive use 
of analytic functions of external variables, the exclusive use of a CNM constructed from white light curve residuals, and their combination 
(see Sect. \ref{sec:step2} for details). We found that the noise structures introduced by the LADC inhomogeneities can be approximated by a combination 
of analytic functions of the parallactic angle, but not perfectly so because this approach lacks accuracy in describing the real signal induced by irregular 
``spots'' on the LADC surfaces. This is reflected by the fact that this approach yields the worst fit to the transit light curves, with residual RMS 
values of 734, 784, and 1132~ppm for the three dates, and large uncertainties on the derived transmission spectra. In addition, transmission spectra 
from different dates show different mean levels, and light curves requiring complicated baseline models (the $\lambda < 788$~nm light curves for 05 December 2012
and 14 January 2013) are offset from the rest of the spectra of each date (top panel in Figure \ref{fig:ssep}). Our second approach, calculating the white 
photometric residuals for each subset of light curves showing similar noise structures, provides a better fit to the data with residual RMS values of 635, 697, 
and 862~ppm, which models the short-timescale structures very well. The resulting transmission spectra show drastically reduced error bars, 85, 25, and 50\% of 
those from the no-CNM analysis for 05 December 2012, 14 January 2013 and 07 February 2013, respectively.  However, at the same time the 
spectra from the three dates disagree substantially, and the 14 January 2013 data show a large trend in the transmission spectrum that is not reproduced in the other data sets.
This trend is most likely a result of sub-optimal modeling of large-scale trends across the light curves, because their amplitudes are 
chromatic, increasing for longer wavelengths. A comparison of the trend amplitude (calculated through the overall pre- and post-transit flux offset) 
with the inferred transmission spectrum shows a very clear (Pearson coefficient of 0.97) correlation for the $\lambda > 788$~nm light curves 
(Fig. \ref{fig:ofc}). 
Finally, we obtained consistent results from all three transits by using a combination of both methods: using low-order polynomials to 
model trends, and the CNM to account for short-timescale variations. Here, the correlation between the slope amplitude and the transmission
spectrum of 14 January 2013 is removed ($p=-0.009$). This approach also yields the best residual RMS values of 635, 697, 
and 862~ppm. Based on this fact and on the excellent agreement of the three dates (average (maximal) disagreement of two measurements
at the same wavelength is 0.5 (1.8) $\sigma$), we find the combined approach to be reliable.

\subsection{Updated system parameters}
We performed joint MCMC analyses of all available data to redetermine the system parameters taking into account the dilution of 
the target flux from the nearby source.
This was done by performing a global analysis using the white FORS2 light curves (as in Sect. \ref{sec:step1}, together with the broadband
data available).  The refined parameters are listed in Table \ref{tab:res} and agree very well (below 1~$\sigma$ for all but $\rho_p$, which 
differs by 1.2~$\sigma$) with those published in \citet{Lendl12}. As a result of the correction for contamination from the newly resolved companion, 
we find a slightly larger the planetary radius ($1.198_{-0.045}^{+0.047}$~{\Rjup} instead of $1.115\pm0.047$~{\Rjup}) and a lower density 
($0.229 \pm 0.016$~{\rhojup} instead of $0.273^{+0.030}_{-0.026}$~{\rhojup}).

\begin{table}
\caption{\label{tab:res} Median values and the 1-$\sigma$ errors of the marginalized posterior PDF 
obtained from the global MCMC analysis.}
\begin{tabular}{ll}       
\hline\hline 
 \multicolumn{2}{l}{Jump parameters} \T  \\
\hline
Stellar metallicity, [Fe/H] [dex] & $-0.23 \pm 0.072$ \T \\ 
Stellar effective temp., $\mathrm{T_{eff}}$ [K] & $5602 \pm 160$ \\ 
Transit depth, $dF$ &  $0.01345 \pm 0.00017$\\ 
Impact parameter, $b'$ [{\Rstellar}] & $0.7704_{-0.0077}^{+0.0072}$\\
Transit duration, $T_{14}$ [d] & $0.08918 \pm 0.00062 $\\
Time of midtransit, $\mathit{T_0}$ $\mathrm{[BJD_{tdb}]}$ & $6267.68389 \pm 0.00013$ \\
Period, $P$ [d] & $2.7817362 \pm 1.4\times10^{-6}$  \\
\hline
\multicolumn{2}{l}{Deduced stellar parameters} \T  \\
\hline
Surface gravity, $\log{g}$ [cgs] & $4.406 \pm 0.019$ \T \\
Mean density, $\rho_{\ast}$ [{\rhosun}] & $0.8934_{-0.036}^{+0.039}$\\
Mass, $M_{\ast}$ [{\Msolar}] & $1.003 \pm 0.10$ \\
Radius, $R_{\ast}$ [{\Rsolar}] & $1.038_{-0.036}^{+0.038}$\\
\hline
\multicolumn{2}{l}{Deduced planet parameters} \T  \\
\hline
Mass, $M_p$ [{\Mjup}] & $0.396 \pm 0.026$  \T \\
Radius, $R_p$ [{\Rjup}] &  $1.198_{-0.045}^{+0.047}$ \\
Semi-major axis, $a$ [au] & $0.03873 \pm 0.0013$ \\
Orbital inclination, $i_p$ [deg] & $84.48 \pm 0.13$ \\
Density, $\rho_{p}$ [{\rhojup}] & $0.229 \pm 0.016$ \\
Surface gravity, $\log{g_{p}}$ [cgs] & $2.853 \pm 0.016$ \\
Equilibrium temp.\tablefootmark{a}, $\mathit{T_{eq}}$ [K] & $1399_{-43}^{+39}$ \\       
 \hline
\multicolumn{2}{l}{Fixed parameters} \T  \\
\hline
Eccentricity, $e$ & 0 \T \\
RV amplitude, $K$ [{\ms}]& $56.8 \pm 2.44$ \\
\hline
\end{tabular}   
\tablefoottext{a}{Assuming an albedo of A=0 and full redistribution from the planet's day to night side, F=1 \T \citep{Seager05}.}
\end{table}

\subsection{Transmission spectrum of WASP-49b}

\begin{figure*}
 \includegraphics[width=\linewidth]{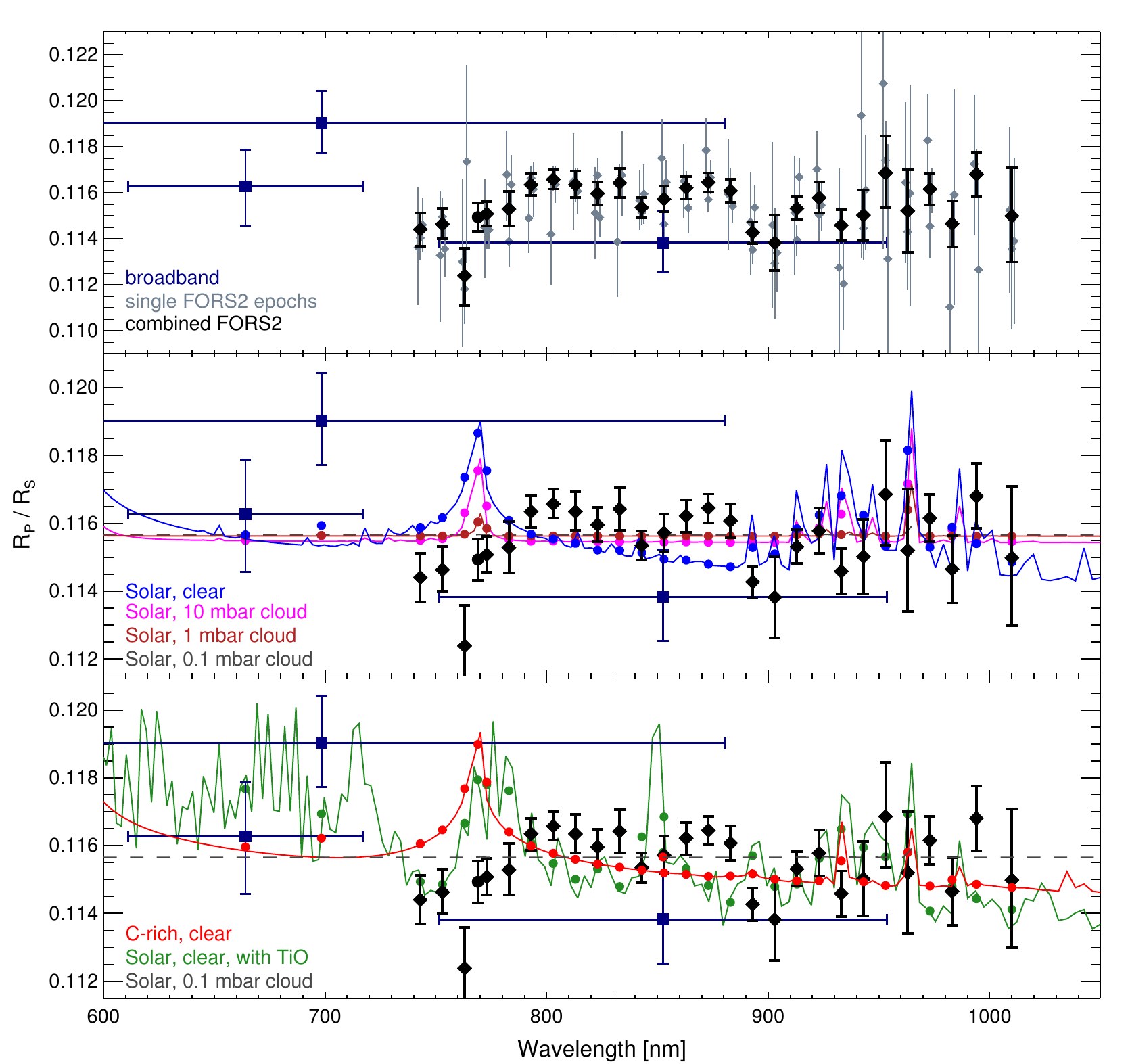}
 \caption{\label{fig:mod}
Transmission spectrum of WASP-49b as observed with FORS2, EulerCam and TRAPPIST compared to atmospheric models. Top: 
the FORS2 results of each separate data set are shown in gray and the results from the combined analysis are shown
as black diamonds. The filled circle represents the data point centered on the K feature, and
the results from broadband light curves are shown as dark blue squares.
Center and bottom: the above transmission spectrum including the combined FORS2 and broadband data, together with predictions 
from atmosphere models. The model atmospheres are a clear solar-composition atmosphere (center panel, blue), a solar-composition
atmosphere with a cloud deck at 10~mbar pressure (central panel, magenta), a solar-composition atmosphere with a cloud deck at
1~mbar pressure (central panel, dark red), a clear C-rich atmosphere (bottom panel, red), 
and a clear solar-composition atmosphere with TiO (bottom panel, green). A flat spectrum obtained from a solar-composition 
atmosphere with a cloud deck at 0.1~mbar pressure is shown as a gray dashed line in the middle and bottom panels. 
The filled circles show the models binned to the observed spectral resolution.}
\end{figure*}

We performed a combined analysis of all FORS2 spectrophotometric light curves
together with the broadband data as described in Sect. \ref{sec:st3}. 
The resulting transmission spectrum of WASP-49b is given in Table \ref{tab:traspec} and shown in Fig. \ref{fig:mod}. 

To interpret the data, we used
physically plausible models of transmission spectra of the planetary
atmosphere. We modeled the transmission spectrum of WASP-49b using the
exoplanetary  atmospheric modeling method of \citet{Madhu09} and 
\citet{Madhu12a}. We considered a plane-parallel atmosphere
at the day-night terminator region that is probed by the transmission
spectrum and computed line-by-line radiative transfer under the
assumption of hydrostatic equilibrium for an assumed temperature
structure and chemical composition. Our plane-parallel atmosphere is
composed of 100 layers, in the pressure range of $10^{-6} - 100$ bar.
We computed the net absorption of the stellar light caused by the
planetary atmosphere as the star light traverses a chord at the
day-night terminator region of the spherical  planet, appropriately
integrated over the annulus.

The model atmosphere includes the major sources of opacity expected in
hot hydrogen-dominated atmospheres, namely, absorption due to alkali
metals (Na and K) and prominent molecules (H$_{2}$O, CO, CH$_{4}$,
CO$_{2}$, C$_2$H$_2$, HCN, TiO), and H$_{2}$-H$_{2}$
collision-induced absorption (CIA) along with gaseous Rayleigh
scattering. The volume mixing ratios of these various species were
chosen assuming chemical equilibrium for different C/O ratios, such as
solar abundance (C/O = 0.5; i.e., oxygen-rich) or carbon-rich (C/O =
1.0; see \citealt{Madhu11a}), but we also explored chemical
disequilibrium solutions if necessitated by the data. In the spectral
range of interest to the current study (i.e., 0.65-1.02 $\mu$m), the
dominant sources of opacity are Na, K, H$_2$O, and TiO. The Na and K
abundances are insensitive to the C/O ratio. However, while H$_2$O and
TiO are abundant in a solar composition atmosphere, they are depleted
by over $\sim$100x for C/O = 1 \citep{Madhu12a}. On the other hand,
we also considered models with an opaque achromatic cloud layer that
effectively obstructs all the spectral features up to a prescribed
cloud altitude; with very high-altitude clouds leading to a
featureless flat spectrum.

We explored the following fiducial model atmospheres with different chemical
compositions to compare with our observed spectra:
\begin{itemize}
 \item a clear Solar-composition atmosphere, without TiO
 \item a clear Solar-composition atmosphere, with TiO
 \item a clear carbon-rich atmosphere ($\mathrm{C/O =1 }$)
 \item Solar-composition atmospheres, without TiO, but with cloud decks at a pressure levels of 0.1, 1, and 10 mbar, respectively.
\end{itemize}

The fits of these models to the observed transmission spectrum are shown in Fig. \ref{fig:mod}.
At our spectral resolution, the model with a cloud deck at 0.1~mbar is essentially identical with a horizontal straight 
line as the clouds obscure all features. For completeness we also compare our data to a constant $R_p/R_\ast$ value.
We compared these models to our data, while compensating for an overall vertical offset between the calculated and 
observed values. The $\chi^2$ values of the available models considering the entire dataset, the FORS2 
dataset alone, or the FORS2 data at $\lambda > 788$~nm alone, are listed in Table \ref{tab:mod}.
 
When considering all available data points, the best fit is obtained by the featureless model of a cloud deck at 0.1~mbar 
pressure, with $\chi^2 = 57.7$. This is nearly identical to the $\chi^2$ of a constant $R_p/R_\ast$ value ($\chi^2 = 57.6$), 
from which a reduced $\chi^2$ of $\chi^2_{red} = 1.86$ is readily calculated, indicating a reasonable fit to the given data. 
A model with a cloud deck at 1~mbar altitude provides a comparably good fit ($\chi^2 = 61.5$), but more complex spectra 
can be excluded. Similar results are obtained if only the FORS2 points are considered, again a cloud decks above 1~mbar 
produce the best fits to the data.
As the largest mismatch between observations and models stems from the wavelength region surrounding the K feature, and
because this region is most affected by strong correlated noise in the light curves, we also tested the 
FORS2 data at wavelengths above 788~nm against the models. We again obtained a good fit for a spectrum with high-altitude 
clouds ($\chi^2=32.4$, $\chi^2_{red}=1.54$), but the carbon-rich model produces a comparably good fit to the data
($\chi^2=30.5$) that is due to its slight slope toward longer wavelengths. A carbon-rich atmosphere would thus still be a 
possibility if the error bars on our short-wavelength measurements are underestimated. 
\begin{table}
 \caption{\label{tab:traspec}
Transmission spectrum of WASP-49b as found by the combined analysis of all \textit{FORS2}, EulerCam, and TRAPPIST data.
} 
\centering
\begin{tabular}{ll}  
\hline\hline 
  Wavelength [nm] & $R_p/{\Rstellar}$ \T \\    
\hline 
  738 - 748  &  $0.1144\pm 0.00072 $  \T       \\
  748 - 758  &  $0.1146_{-0.00063}^{+0.00068}$ \\
  758 - 768  &  $0.1124\pm 0.0013 $            \\
765 - 773 (K)&  $0.1151_{-0.00052}^{+0.00054}$  \\
  768 - 778  &  $0.1153_{-0.00075}^{+0.00077}$ \\
  778 - 788  &  $0.1149\pm 0.00062$            \\
  788 - 798  &  $0.1163_{-0.00048}^{+0.00046}$ \\
  798 - 808  &  $0.1166\pm 0.00043$            \\
  808 - 818  &  $0.1163_{-0.00055}^{+0.00058}$ \\
  818 - 828  &  $0.1160\pm 0.00052 $           \\
  828 - 838  &  $0.1164\pm 0.00063 $           \\
  838 - 848  &  $0.1154\pm 0.00044 $           \\
  848 - 858  &  $0.1157\pm 0.00056 $           \\
  858 - 868  &  $0.1162_{-0.00049}^{+0.00047}$ \\
  868 - 878  &  $0.1165_{-0.00043}^{+0.00041}$ \\
  878 - 888  &  $0.1161\pm 0.00051 $           \\
  888 - 898  &  $0.1143\pm 0.00048 $           \\
  898 - 908  &  $0.1138\pm 0.0012 $            \\
  908 - 918  &  $0.1153\pm 0.00051 $           \\
  918 - 928  &  $0.1158\pm 0.00068 $           \\
  928 - 938  &  $0.1146\pm 0.00067 $           \\
  938 - 948  &  $0.1150\pm 0.0011 $            \\
  948 - 958  &  $0.1169\pm 0.0016 $            \\
  958 - 968  &  $0.1152\pm 0.0018 $            \\
  968 - 978  &  $0.1162\pm 0.00070$            \\
  978 - 988  &  $0.1147\pm 0.0010$             \\
  988 - 1000 & $0.1168\pm 0.00096 $            \\
 1000 - 1020 & $0.1150\pm 0.0021 $             \\
611 - 717\tablefootmark{a} & $0.1163\pm 0.0017$ \\
516 - 880\tablefootmark{b} & $0.1190\pm 0.0014$ \\
751 - 953\tablefootmark{c} & $0.1138\pm 0.0014$ \\  
\hline
\end{tabular}
 \newline
 \tablefoottext{a}{\textit{r'-Gunn} filter}\T,
 \tablefoottext{b}{\textit{NGTS} filter},
 \tablefoottext{c}{\textit{Ic+z'-Gunn} filter}
\end{table}
\begin{table}
 \caption{\label{tab:mod}$\chi^2$ values calculated from the data and the various atmosphere models for WASP-49b. 
} 
\centering
\begin{tabular}{lccc}  
\hline\hline 
Model & All & FORS2 & FORS2, $\lambda>788$~nm \T \\
\hline
\multicolumn{4}{l}{$\chi^2$ values} \T \\
\hline
Carbon-rich         & 158.5 & 152.3 & 30.5 \T \\                
Solar (no TiO)      & 169.8 & 162.4 & 63.6 \\                
Solar (with TiO)    & 173.2 & 167.9 & 84.4 \\                
Solar (10~mbar cloud)  & 95.8  & 87.4  & 42.4 \\                
Solar (1~mbar cloud) & 61.5 & 53.2 & 33.3 \\                
Solar (0.1~mbar cloud) & 57.7 & 49.5 & 32.4 \\                
Constant            & 57.6  & 49.3  & 32.4 \\ 
\hline 
\multicolumn{4}{l}{$\chi^2_{red}$ values} \T \\
\hline              
Constant  & 1.86  & 1.83  & 1.54 \T \\                
\hline
\end{tabular}
\end{table}

\section{Conclusions}
\label{sec:W49con}
We have obtained a transmission spectrum of WASP-49b based on VLT/FORS2 observations of three planetary transits. The FORS2 data are 
affected by considerable systematic noise due to LADC inhomogeneities, but this noise is limited for observations where
the LADC prism separation was set to a minimum and kept constant throughout the observation. We found consistent results from 
all three dates only when we applied a common noise model for light curve sets showing similar correlated noise together with low-order
polynomial baselines to model each light curve's large-scale trends individually. We therefore warn against the ``blind'' use of 
white light curve residuals alone to model spectrophotometric light curves that are affected with substantial correlated noise.

Using these data, we also updated the system parameters by taking contamination from a newly discovered nearby star into account. Our data
agree with the previously published values while favoring a slightly larger planetary radius ($1.198_{-0.045}^{+0.047}$~{\Rjup} 
instead of $1.115\pm0.047$~{\Rjup}) and hence a lower planetary bulk density ($0.229 \pm 0.016$~{\rhojup} instead of $0.273^{+0.030}_{-0.026}$~{\rhojup}).
The transmission spectra we obtain from the three epochs agree well with each other, demonstrating the instrumental 
stability and usefulness of FORS2 for high-precision spectrophotometry even in the presence of LADC-induced correlated noise. 

We found that the transmision spectrum of WASP-49b is best fit by models with muted spectral features, 
such as expected in the presence of opaque high-altitude clouds or hazes. A carbon-rich atmosphere provides a comparable fit only 
when data at  \mbox{$\lambda < 788$~nm} are removed from the analysis. Solar-composition atmospheres, both with and without TiO are a poor match to the data.
We conclude that WASP-49b most likely has clouds or hazes at pressure levels of 1~mbar or less.

\begin{acknowledgements}

We would like to thank our referee, Drake Deming, for insightful comments that improved the 
quality of this manuscript, and Amaury Triaud for helpful scientific discussions.
This work was supported by the European Research 
Council through the European Union's Seventh Framework Programme 
(FP7/2007-2013)/ERC grant agreement number 336480. 
TRAPPIST is funded by the Belgian Fund for Scientific  
Research (Fond National de la Recherche Scientifique, FNRS) under the  
grant FRFC 2.5.594.09.F, with the participation of the Swiss National  
Science Fundation (SNF). L. Delrez acknowledges the support of the F.R.I.A. 
fund of the FNRS. M. Gillon and E. Jehin are FNRS Research Associates.

\end{acknowledgements}

\bibliographystyle{aa}
\bibliography{./bbl}

\end{document}